\documentclass{article}
\usepackage{tikz}
\usetikzlibrary{tikzmark}
\AtBeginDocument{%
  \providecommand\BibTeX{{%
    \normalfont B\kern-0.5em{\scshape i\kern-0.25em b}\kern-0.8em\TeX}}}

\usepackage[algo2e,vlined,algoruled]{algorithm2e}

\usepackage{amsmath,amssymb,amsfonts,xcolor,graphicx}
\usepackage{doi}
\usepackage{amsthm}
\usepackage{listings}
\usepackage{bbold}
\usepackage{bm}

\usepackage{subfig}
\usepackage{layouts}

\newcommand{\rvect}[1]{\begin{bmatrix} #1 \end{bmatrix}}

\DeclareMathOperator{\Tr}{Tr}

\input{Qcircuit}

\usepackage{thmtools}
\usepackage{nameref}
\usepackage[nameinlink]{cleveref}
\crefname{lemma}{lemma}{lemmas}
\Crefname{lemma}{Lemma}{Lemmas}
\crefname{assumption}{assumption}{assumptions}
\Crefname{assumption}{Assumption}{Assumptions}
\crefname{algocf}{alg.}{algs.}
\Crefname{algocf}{Algorithm}{Algorithms}
\Crefname{algocfline}{Line}{Lines}

\usepackage[sort&compress,square,comma,numbers]{natbib}
\usepackage{fullpage,authblk}
\hypersetup{breaklinks=true, colorlinks=true, linkcolor={blue!90!black}, citecolor={blue!90!black}, urlcolor={blue!90!black}}

\usepackage[normalem]{ulem} %

\begin{document}

\title{Hardware-Conscious Optimization of the Quantum Toffoli Gate}

\author[a]{Max Aksel Bowman}

\author[b]{Pranav Gokhale}

\author[c]{Jeffrey Larson}
\author[c]{Ji Liu}

\author[d]{Martin Suchara}

\affil[a]{Rice University, {\tt mab31@rice.edu}}
\affil[b]{Super.tech, a division of ColdQuanta, {\tt pranav.gokhale@coldquanta.com}}
\affil[c]{Argonne National Laboratory, {\tt jmlarson@anl.gov}, {\tt ji.liu@anl.gov}}
\affil[d]{Amazon, {\tt sucharam@amazon.com}}
\date{}                     %
\renewcommand\Affilfont{\itshape\small}

\maketitle
\begin{abstract}

While quantum computing holds great potential in combinatorial optimization, electronic structure calculation, and number theory, the current era of quantum computing is limited by noisy hardware. Many quantum compilation approaches can mitigate the effects of imperfect hardware by optimizing quantum circuits for objectives such as critical path length. Few approaches consider quantum circuits in terms of the set of vendor-calibrated operations (i.e., native gates) available on target hardware. This manuscript expands the analytical and numerical approaches for optimizing quantum circuits at this abstraction level. We present a procedure for combining the strengths of analytical native gate-level optimization with numerical optimization. Although we focus on optimizing Toffoli gates on the IBMQ native gate set, the methods presented are generalizable to any gate and superconducting qubit architecture. Our optimized Toffoli gate implementation demonstrates an $18\%$ reduction in infidelity compared with the canonical implementation as benchmarked on IBM Jakarta with quantum process tomography. Assuming the inclusion of multi-qubit cross-resonance (MCR) gates in the IBMQ native gate set, we produce Toffoli implementations with only six multi-qubit gates, a $25\%$ reduction from the canonical eight multi-qubit implementations for linearly connected qubits.

\end{abstract}

\section{Introduction}
Over the past decade, quantum hardware has improved enough to
physically realize quantum computers with tens of noisy
qubits~\cite{preskill2021quantum, Arute2019, wright2019benchmarking,
jurcevic2021demonstration}. Coupled with the creation of quantum software
development kits such as IBM's Qiskit~\cite{Qiskit} and Google's
Cirq~\cite{Cirq}, the current era of quantum computing has seen unprecedented
researcher access to state-of-the-art quantum computers. With this broader
access has come an accelerated demand for mitigating  physical limitations that
confine modern quantum circuit lengths to depths well under 100 multi-qubit
gates. While several physics-based techniques for noise mitigation exist, such
as dynamical decoupling~\cite{dynamicdecoupling2010, das2021adapt} and echoed
gates~\cite{sundaresan2020reducing, sheldon2016procedure}, much success has
been found in developing quantum  compilation strategies that yield circuits
with smaller depths and fewer necessary operations.

A plethora of modern approaches to quantum compilation aim to decrease compiled circuit length and increase algorithm fidelity. While many of these approaches leverage classical compilation techniques such as temporal planners~\cite{temporalplanners2018}, list schedulers~\cite{lao2021}, and DAG circuit representations~\cite{li2019tackling}, physical limitations inherent in quantum computation have necessitated the fusion of these techniques with novel hardware-conscious methods that transcend traditional compiler theory. A particularly profitable class of these methods has tackled the problem of efficiently mapping quantum circuits to underlying qubit
connectivity~\cite{tan2021optimal, tannu2019ensemble, patel2020ureqa,
li2019tackling,
tannu2019not, Duckering2021, finigan2018qubit}. Another class of methods
provides quantum compilers with accurate noise models of their target hardware
in order to improve program success~\cite{murali2020software,
tannu2019mitigating}.

The success of the preceding methods lends much support to taking a
hardware-conscious approach to quantum algorithm optimization. Nevertheless, few current approaches consider optimization at the native gate level, a level of abstraction below that of standard quantum algorithm specification in
which programs are described in terms of the quantum computer's native gate
set, namely, the set of operations explicitly calibrated by the hardware
vendor. Previous work has explored the optimization of gates such as
the
SWAP gate~\cite{gokhale2021faster} and ZZ interaction~\cite{gokhale2020optimized} at
the native gate level, and only recently have researchers turned their
attention to optimizing slightly longer operations such as the Toffoli gate at
this level of abstraction~\cite{oomura2021design}.

The Toffoli gate is a fundamental three-qubit, multi-control operation that enacts a Pauli-X gate (the quantum equivalent of a classical NOT operation) on its target qubit provided that its two control qubits are in the $\ket{1}$ state~\cite{ToffoliReversibleComputing, shende2008cnotcost}. The Toffoli gate is used in many important  quantum computing applications, including Shor's algorithm~\cite{Shor_1997}, Grover's algorithm~\cite{Grover1996, Figgatt2017}, Takahashi adders~\cite{takahashi2009quantum},  and quantum error correction~\cite{Reed_2012, crow2016improved}.
Many of these applications enable some of the most exciting possibilities offered by quantum computing, underscoring the importance of the Toffoli gate.

Optimizing the Toffoli gate is crucial to  improving quantum circuit compilation.  For example, it has been shown that delayed hardware-aware decomposition of the Toffoli gate to the native gate set of target hardware  substantially improves compilation results~\cite{Duckering2021}.  Furthermore, Amdahl's law~\cite{Hill2008} suggests that small optimizations of the Toffoli gate can yield exponential fidelity improvements (and hence error reductions) for the many quantum algorithms that leverage it, making its native gate-level optimization worthwhile to explore.

In this paper, we optimize the Toffoli gate for the IBMQ native gate set via a combination of existing and novel native gate-level optimization techniques. It is important to note that although we focus on the Toffoli gate, the methods developed in this paper are readily generalizable to other quantum gates and circuits. The major contributions of this paper are listed below.

\begin{itemize}
    \item A novel native gate-level optimization technique termed "side-effect sandwiching" that significantly reduces quantum circuit critical path length
    \item An optimized Toffoli gate for IBMQ machines with $18$\% lower infidelity than the current standard as benchmarked on IBMQ Jakarta
    \item A method of parameterizing the set of all $m$-qubit quantum circuits with $n$ multi-qubit operations with $3m(n+1)$ real numbers
    \item Numerical quantum circuit optimization methodology complementary to existing analytical native gate-level techniques
\end{itemize}

This paper is organized as follows. \Cref{sec:background} contains background information on quantum computing and a review of existing native gate-level optimization techniques (also referred to as "a priori techniques").
\Cref{sec:apriori} introduces side-effect sandwiching, a is a novel technique. Native gate-level optimization is used to produce an optimal Toffoli gate for the IBMQ native gate set which is then characterized via quantum process tomography.
\Cref{sec:numerical} develops a methodology for the numerical discovery of novel quantum circuit implementations of a provided unitary matrix as well as how numerical techniques can aid analytical a priori techniques.
\Cref{sec:applications} demonstrates the optimality of the Toffoli gate circuit presented in \Cref{sec:apriori} by applying it to quantum compilation.
Finally, we conclude and identify future research directions in \Cref{sec:conclusion}.

\section{Background}\label{sec:background}
In this section we briefly introduce quantum circuits, quantum computer
native gate sets, the multi-target cross-resonance gate, current
Toffoli gate implementations for superconducting qubits, and some native gate-level optimization techniques. An exploration of the limitations of existing hardware and qubit connectivity is also provided. Moreover, notation specific to this paper is introduced.

\subsection{Qubits and Quantum Circuits}
The most fundamental unit of quantum information is the
qubit. The qubit is represented mathematically as a linear combination $\ket{\psi} = \alpha \ket{0} + \beta
\ket{1}$ of basis vectors $\ket{0}$ and $\ket{1}$, where $\alpha, \beta \in
\mathbb{C}$ and $|\alpha|^2+|\beta|^2=1$. Note that this expression indicates
that a qubit is not limited to existing in either the $\ket{0}$ or $\ket{1}$
state. Instead, it may also occupy a superposition of these two states. A
single qubit state $\ket{\psi}$ can be represented graphically as a point on
the surface of a Bloch sphere.

Two qubits are said to be entangled if the state of the two-qubit system cannot
be described as the tensor product of individual qubit states. Entangled
states, while much more difficult to visualize and work with, are useful for a
variety of applications including super-dense coding and state teleportation.
This phenomenon, along with superposition, enables quantum computers to solve
certain problems asymptotically faster than classical computers can.

In order to perform a certain computation on qubits, a series of quantum
operations or gates performing this computation must be specified. Quantum
circuits mirror classical digital logic circuits in that they are specified as
a sequence of quantum gates to execute on some subset of available qubits.
Quantum gates are represented algebraically by unitary matrices that describe
how they map input quantum states to output quantum states and theoretically
operate on any number of qubits. A quantum algorithm can be specified as a
quantum circuit along with a procedure for interpreting the final quantum state
resulting from the computation.

The lowest level of abstraction for specifying quantum circuits on
superconducting qubits is the pulse schedule. Superconducting qubits are
controlled with microwave pulses created by arbitrary waveform generators. Each
native gate in the native gate set of a quantum computer has an associated
pulse schedule specifying the amplitude of the microwave pulse on a given qubit
at any moment in time. In practice, the time axis is discretized. Pulse-level
access to quantum computers gives the user more control over how operations are
executed. In this paper we use pulse-level control in order to engineer new
native gates and realize our native gate-level optimizations on modern
quantum hardware.

\subsection{Quantum Hardware Limitations}
Similar to classical computers, quantum computers are limited by the way their computing model is physically implemented. Different quantum machines have different vendor-calibrated operations (i.e., native gates) available, which presents a challenge to developing quantum compilers. These constraints can sometimes be relaxed by pulse-engineering custom operations, effectively augmenting the native gate set of target hardware. This limitation is covered in \Cref{native_gate_set_intro}.  Different quantum computers also have different levels of qubit connectivity, which has a significant impact on the implementation of quantum circuits on quantum hardware; the majority of the length of NISQ-era circuits involve swapping information between qubits to satisfy hardware connectivity constraints. This constraint is covered in \Cref{qubit_routing}.

\subsubsection{Native Gate Sets}\label{native_gate_set_intro}
Quantum algorithm design often proceeds in a relatively hardware-agnostic
fashion, with quantum circuits built from a set of abstract, mathematically
convenient gates that must be later transpiled to available vendor-calibrated
operations and mapped onto the underlying qubit architecture of the
device~\cite{cowtan_et_al:LIPIcs:2019:10397}. Consideration of the natural
limitations of the underlying hardware during algorithm design yields several
opportunities for optimization. The native gate set of a quantum computer is
defined as the set of operations specifically calibrated by the hardware vendor
to perform a targeted unitary operation. This set of operations is
analogous to the instruction set in classical computing. Recent results in the
quantum compiler community have indicated great opportunities in the optimization
of quantum circuits at the native gate set-level~\cite{gokhale2020optimized}. In
the case of IBMQ hardware, the native gate set consists of $\sqrt{X}$, $X$, the
cross-resonance operator, and $R_z(\theta)$ for arbitrary $\theta$. The
single-qubit gates $\sqrt{X}$ and $X$ enact $90^\circ$ and $180^\circ$
rotations of a qubit state about the $x$-axis of the Bloch sphere, while the
$R_z(\theta)$ gate enacts a $\theta$ angle rotation of a qubit state about the
$z$-axis of the Bloch sphere. The cross-resonance operator enables interaction
between pairs of qubits and effectively performs a controlled-$R_x$ gate. When
the rotation angle is calibrated to $pi$, the cross-resonance effectively
executes a CNOT gate (up to single-qubit gates). The cross-resonance operator
is generally implemented as two half-cross-resonance pulses of opposite polarity so as to mitigate error. The details of this procedure are below the
native gate set abstraction level; only the notion that there are two polarities for half-cross-resonance pulses is necessary for our discussion of native gate-level optimization.

Prior research indicates that augmenting native gate sets with
pulse-level hardware control (such as OpenPulse for IBMQ hardware)
leads to fidelity improvements in quantum gate
implementations~\cite{gokhale2021faster}, including the Toffoli
gate~\cite{oomura2021design}. In this paper we consider augmenting the IBMQ
native gate set to include the multi-target cross-resonance gate, which executes the equivalent of a cross-resonance gate between a target qubit and every target qubit. The single-target cross-resonance gate is normally implemented by driving its control qubit at the resonant frequency of its target qubit with an arbitrary waveform generator. The multi-target cross-resonance gate is implemented the same way, except that its control qubit is driven with the superposition of microwave pulses with resonant frequencies of each of its target qubits. The multi-target cross-resonance gate was chosen for two reasons. First, its operation is easily understood in terms of a familiar gate (i.e., the single-qubit cross-resonance gate), which makes analytical optimizations more tractable. Second, it has already been demonstrated to asymptotically improve the length of several high-impact quantum algorithms~\cite{gokhale2020quantum}.

\subsubsection{Qubit Routing}\label{qubit_routing}
\begin{figure}[h]
    \centering
    \includegraphics[width=0.2\textwidth]{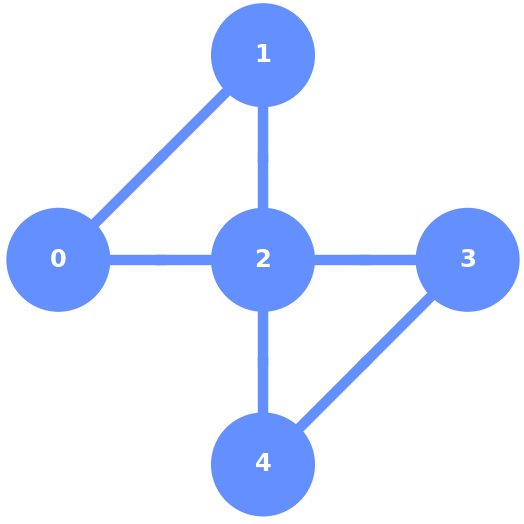}
    \hspace{1in}
    \includegraphics[width=0.2\textwidth]{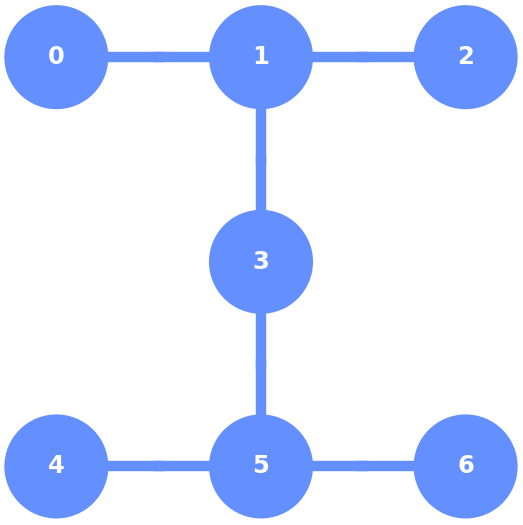}
    \caption{The connectivity graph of the IBMQ Yorktown qubit architecture (left) features two fully connected three-qubit subgraphs, while the IBMQ Jakarta qubit architecture (right) features only linearly connected three-qubit subgraphs.}
    \label{fig:connectivity}
\end{figure}

When optimizing quantum algorithms for a target machine, the underlying qubit connectivity must be considered. Cross-resonance gates
can be executed only on certain pairs of qubits in most superconducting qubit
architectures. If a cross-resonance gate can be executed between two
qubits, they are said to be "connected," and a graph with qubits as nodes can
be constructed that shows the qubit connectivity of a quantum computer. In this
paper, since we are optimizing a three-qubit gate, we focus on the connectivity
of subgraphs of three nodes. We define a three-qubit subgraph as fully
connected if every qubit in the subgraph is connected to every other qubit in
the subgraph. We define a three-qubit subgraph as being linearly connected if the
subgraph is connected and contains one qubit with degree two and two qubits with degree one.
\Cref{fig:connectivity} shows qubit connectivity graphs of two IBMQ
systems; note that the left graph contains two fully connected three-qubit
subgraphs, while the right graph contains only linearly connected three-qubit
subgraphs.

Compiling quantum circuits to respect the underlying qubit connectivity of a
target system is termed the qubit routing problem, and it represents an
important challenge in quantum compilation. The current method of
solving the qubit routing problem is inserting SWAP operations that switch the
quantum states represented by two connected qubits. If a cross-resonance
gate is to be performed between two unconnected qubits, the states on
these qubits can be swapped to adjacent qubits that can execute the operation.
The importance of qubit routing is demonstrated by the effort that has been
devoted to reducing the number of required SWAP operations to run algorithms on
sparsely connected systems by the quantum compilation community. On 
most superconducting qubit systems (especially systems with over ten
qubits), SWAP operations account for most of the compiled circuit and
indicate the limitations of the underlying qubit
connectivity.

Different connectivities have different strengths. On superconducting qubit
architectures, operations run on fully connected qubit subgraphs require fewer
SWAP operations in their implementations but are also more prone to crosstalk
error. Linearly connected qubit subgraphs, on the other hand, require more SWAP
operations but are also less prone to crosstalk error; in fact, quantum
computer providers such as IBMQ are increasingly using less-connected
architectures to improve qubit fidelity~\cite{Ibm2021}. Therefore, in this paper, we focus on developing optimizations for linearly connected qubits.

\subsection{Notation}
In this section, we introduce some notation. To
denote the unitary of the Toffoli gate, we use the symbol $U_{CCX}$. To denote a $90^\circ$ rotation around the $x-axis$ of the Bloch sphere, we write $\sqrt{X}$. To denote a negative $90^\circ$ rotation around the $x$-axis of the Bloch sphere, we write
$R_z(180^\circ)\sqrt{X}R_z(180^\circ)$.

$R_z$ rotations are not physically
executed and can be kept track of
in software. Therefore, they are unimportant when measuring circuit parameters
such as circuit length and total degrees of rotation. To emphasize this point,
we denote $z$-axis rotations of various common angles with small arrows as shown
in \Cref{table:zrot}.

\begin{table}[h]
\centering
  \caption{Circuit symbols (bottom) for representing the $R_z$ rotation of various angles in radians (top).}
  \label{table:zrot}
\begin{tabular}{cccccccc}
$\frac{\pi}{4}$ & $-\frac{\pi}{4}$ & $\frac{\pi}{2}$ & $-\frac{\pi}{2}$ & $\frac{3\pi}{4}$ & $-\frac{3\pi}{4}$ & $\pi$ & $-\pi$ \\
$\nearrow$ & $\nwarrow$ & $\curvearrowright$ & $\curvearrowleft$ & $\searrow$ & $\swarrow$ & $\circlearrowright$ & $\circlearrowleft$ \\ 
\end{tabular}
\end{table}

\subsection{Existing Toffoli Gate Implementations}
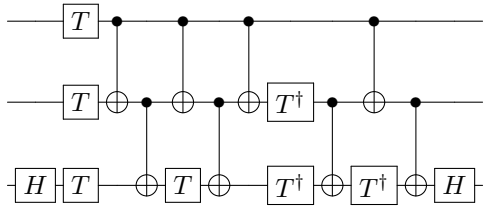
\begin{figure}[h]
    \[
    \Qcircuit @C=0.3em @R=1.7em {
    & \qw & \gate{T} & \ctrl{1} & \qw & \ctrl{1} & \qw & \ctrl{1} & \qw & \qw & \ctrl{1} & \qw & \qw & \qw & \\
    & \qw & \gate{T} & \targ & \ctrl{1} & \targ & \ctrl{1} & \targ & \gate{T^\dag} & \ctrl{1} & \targ & \ctrl{1} & \qw & \qw &\\
    & \gate{H} & \gate{T} & \qw & \targ & \gate{T} & \targ & \qw & \gate{T^\dag} & \targ & \gate{T^\dag} & \targ & \gate{H} & \qw &\\
    }
    \]
    \caption{Canonical implementation of a Toffoli gate designed for linear qubit architectures.}
    \label{fig:lc_toffoli}
\end{figure}

There are several existing implementations of the Toffoli gate for superconducting architectures. One of these is shown in \Cref{fig:lc_toffoli}. We refer to this implementation~\cite{Nielsen2009} as the "canonical linear Toffoli gate" because it is among the oldest and most commonly used quantum circuits for implementing the Toffoli gate on linearly connected qubits. Previous work has explored optimizing the Toffoli gate at the pulse level~\cite{oomura2021design}, but the produced optimization introduces a side-effect SWAP gate on the control qubits, which may not always be desired.

\subsection{Native Gate-Level Optimization}
Although consideration of quantum circuits at the native gate
level is a generally novel approach, a number of optimization techniques at this abstraction level have been explored in prior literature. We use the SuperstaQ platform \cite{superstaq} to apply three existing native gate-level optimization techniques to the Toffoli gate, initially developed to
optimize the quantum SWAP gate~\cite{gokhale2021faster}. 
Native gate-level optimization starts with a standard gate-level
specification of a quantum circuit. The quantum circuit to be optimized is then
translated into the native gate set of the target quantum hardware. Next, the
following three techniques can be used to reduce circuit depth and active
rotation.

\subsubsection{Cross-Gate Pulse Cancellation}\label{sec:cgpc}
Cross-gate pulse cancellation is a technique in which a sequence of native gates is compressed into a shorter sequence of native gates (including possibly the NO-OP identity); For example, if during optimization the native gate sequence
$\sqrt{X}X\sqrt{X}$ is encountered (the $\sqrt{X}$ may itself be part of a preceding sequence of gates), it can be eliminated because it is
equivalent to a $360^\circ$ (and $0^\circ$) rotation around the $x$-axis of the Bloch sphere. In this case, cross-gate pulse cancellation would save
$360^\circ$ of active rotation and potentially result in circuit length
reduction if one or more of the native gates were part of the critical path
length of the quantum circuit.

\subsubsection{Cross-Resonance Commutation}
\begin{figure}[h]
\[\begin{array}{l}
    \Qcircuit @C=0.3em @R=.7em {
    & \ctrl{1} & \qw & \\
    & \gate{+} & \qw & \\
    } =
    \Qcircuit @C=0.3em @R=.7em {
    & \ctrlo{1} & \qw & \ctrl{1} & \qw & \gate{X} & \ctrl{1} & \qw & \ctrlo{1} & \qw & \\
    & \gate{X^{\frac{1}{4}}} & \circlearrowleft \qw & \gate{X^{\frac{1}{4}}} & \qw & \circlearrowright \qw & \gate{X^{\frac{1}{4}}} & \circlearrowleft \qw & \gate{X^{\frac{1}{4}}} & \circlearrowright \qw & \\
    }
\end{array}\]
\caption{Model of the positive polarity cross-resonance operator in terms of controlled x-rotations.}
\label{fig:cr_commutation}
\end{figure}
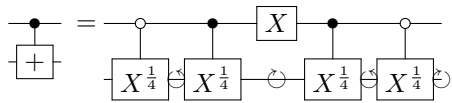

Gates enacting rotations around the $x$-axis of the Bloch sphere commute. This fact becomes particularly relevant when considering the controlled x-rotation model of the cross-resonance operator. \Cref{fig:cr_commutation} shows a conceptual model of the positive polarity cross-resonance operator. Note that half-CR pulses can be modeled as controlled $X^{\frac{1}{4}}$ gates. This permits the movement of gates enacting $x$-axis Bloch sphere rotations on the target qubit through the cross-resonance operator. This often results in opportunities for active rotation and circuit length reduction via cross-gate pulse cancellation.

\subsubsection{Cross-Resonance Polarity Swap}
The cross-resonance operator has an associated polarity. Positive polarity cross-resonance operators are
composed first of a positive half-CR pulse followed by an $X$ gate on the
control qubit and then a negative half-CR pulse. Negative polarity cross
resonance operators are the same except they first execute a negative half-CR
pulse followed by an $X$ gate and then a positive half-CR pulse. Implementing
a negative polarity cross-resonance operator, however, introduces a side effect
on the control qubit that must be counteracted with an $X$ gate preceding the
operation. This extra $X$ gate can eliminate other gates or gate
sequences via cross-gate pulse cancellation.

\subsubsection{Euler Decomposition}\label{sec:euler}
Another technique that yielded considerable savings in angle rotation
is decomposing the unitary of long single-qubit gate sequences into shorter, equivalent sequences. For example, CR gate commutation and cross-gate
pulse cancellation often yield gate sequences featuring an $R_z$ gate
sandwiched between two $R_x$ gates. Some of these gate sequences occur multiple
times; in many cases, one can find an equivalent gate sequence with fewer
$\sqrt{X}$ and $X$ gates and more degrees of $R_z$ rotation. $R_z$
gates do not represent physical rotations performed on IBMQ systems but are
viewed instead as a change in the Pauli reference frame \cite{mckay2017efficient}. Therefore, $R_z$ gates
are better thought of as book-keeping operations rather than physical
processes. Therefore, this method effectively replaces physical quantum
computer operations with "virtual" operations that do not need to be physically
implemented. By reducing the number of physical operations required to perform
a quantum circuit, fewer places exist for introducing errors.

\section{Analytical A Priori Approach}\label{sec:apriori}
In this section, we apply several native gate-level optimization techniques to the Toffoli gate implementation in \Cref{fig:lc_toffoli}. We also present cross-resonance gate sandwiching, a novel technique that can significantly reduce the length of quantum circuits. All of these methods are termed "a priori" because they operate on an already existing quantum circuit.

\subsection{Cross-Resonance Sandwiching}
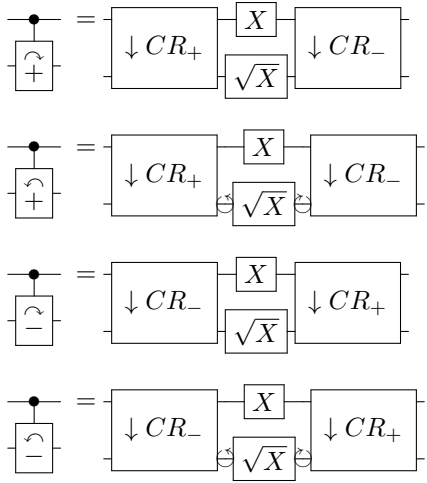
\begin{figure}
\[\begin{array}{l}
    \Qcircuit @C=0.3em @R=.7em {
    & \ctrl{1} & \qw & \\
    & \gate{\overset{\curvearrowright}{+}} & \qw & \\
    } =
    \Qcircuit @C=0.3em @R=.7em {
    & \multigate{1}{\downarrow CR_{+}} & \gate{X} & \multigate{1}{\downarrow CR_{-}} & \qw & \\
    & \ghost{\downarrow CR_{-}} & \gate{\sqrt{X}} & \ghost{\downarrow CR_{-}} & \qw & \\
    }\\ \\
    \Qcircuit @C=0.3em @R=.7em {
    & \ctrl{1} & \qw & \\
    & \gate{\overset{\curvearrowleft}{+}} & \qw & \\
    } =
    \Qcircuit @C=0.3em @R=.7em {
    & \multigate{1}{\downarrow CR_{+}} & \qw & \gate{X} & \qw & \multigate{1}{\downarrow CR_{-}} & \qw & \\
    & \ghost{\downarrow CR_{+}} & \circlearrowleft \qw & \gate{\sqrt{X}} & \circlearrowright \qw & \ghost{\downarrow CR_{-}} & \qw & \\
    } \\ \\
    \Qcircuit @C=0.3em @R=.7em {
    & \ctrl{1} & \qw & \\
    & \gate{\overset{\curvearrowright}{-}} & \qw & \\
    } =
    \Qcircuit @C=0.3em @R=.7em {
    & \multigate{1}{\downarrow CR_{-}} & \gate{X} & \multigate{1}{\downarrow CR_{+}} & \qw & \\
    & \ghost{\downarrow CR_{-}} & \gate{\sqrt{X}} & \ghost{\downarrow CR_{+}} & \qw & \\
    }\\ \\
    \Qcircuit @C=0.3em @R=.7em {
    & \ctrl{1} & \qw & \\
    & \gate{\overset{\curvearrowleft}{-}} & \qw & \\
    } =
    \Qcircuit @C=0.3em @R=.7em {
    & \multigate{1}{\downarrow CR_{-}} & \qw & \gate{X} & \qw & \multigate{1}{\downarrow CR_{+}} & \qw & \\
    & \ghost{\downarrow CR_{-}} & \circlearrowleft \qw & \gate{\sqrt{X}} & \circlearrowright \qw & \ghost{\downarrow CR_{+}} & \qw & \\
    }
\end{array}\]
\caption{Definitions of sandwiched cross-resonance operations. The sign
indicates the polarity of the cross-resonance operations, and the direction of
the arrow represents whether a $90^\circ$ or $-90^\circ$ side-effect rotation
should be applied to the target qubit.}
\label{fig:sandwiched}
\end{figure}

\begin{figure}
    \centering
    \includegraphics[width=0.45\textwidth]{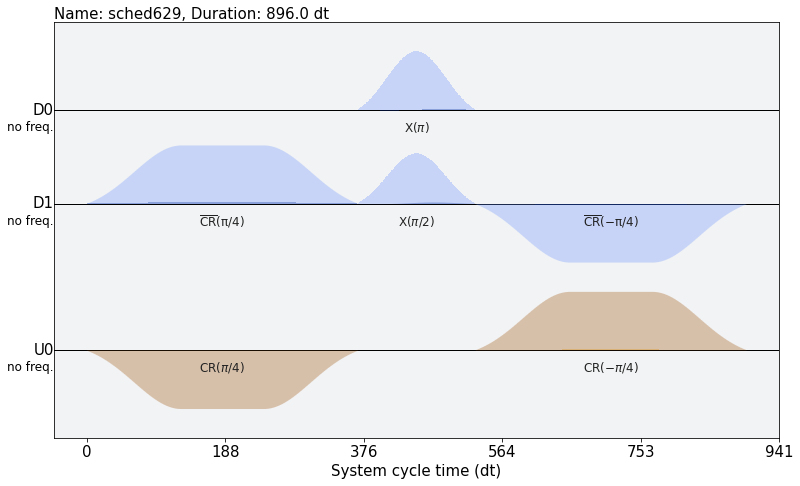}
    \caption{Pulse schedule of a sandwiched positive polarity cross-resonance
    gate enacting a $\sqrt{X}$ side-effect on its target qubit.}
    \label{fig:ps}
\end{figure}

In addition to leveraging the prior native gate-level work covered in \Crefrange{sec:cgpc}{sec:euler}, we introduce cross-resonance gate sandwiching, a novel technique that yielded
considerable savings in Toffoli gate implementation length. The
cross-resonance operator is generally implemented in an echoed fashion that
leaves room for a single-qubit gate to be applied to the target
qubit at the same time as an $X$ gate is applied to the control qubit. This
observation is useful because CR gate commutation allows for $\sqrt{X}$ and
$R_z(180^\circ)\sqrt{X}R_z(180^\circ)$ gates to be moved inside cross-resonance
operators. In the context of computer architecture, this technique can be
thought of as pipelining single-qubit operations. Although single-qubit
operations are generally much faster than multi-qubit operations, repeated
sandwiching of bottleneck single-qubit operations into multi-qubit operations
adds up to significant time savings in certain cases. Examples and notations
regarding this technique are shown in \Cref{fig:sandwiched}. 
\Cref{fig:ps} shows the pulse schedule implementation of the cross-resonance
operation with positive polarity and a $\sqrt{X}$ side-effect. This
optimization is enabled only by pulse-level control of the hardware.

\subsection{Results}
\begin{figure}
\centering
\[\begin{array}{c}
    \Qcircuit @C=0.4em @R=1.7em {
    & \searrow \qw & \ctrl{1} & \curvearrowleft \qw & \qw & \ctrl{1} & \curvearrowright \qw & \qw & \ctrl{1} & \curvearrowleft \qw & \qw & \ctrl{1} & \qw & \qw & \qw & \qw & \qw\\
    & \nearrow \qw & \gate{\overset{\curvearrowright}{-}} & \curvearrowright \qw & \ctrl{1} & \gate{\overset{\curvearrowright}{+}} & \curvearrowleft \qw & \ctrl{1} & \gate{\overset{\curvearrowleft}{-}} & \nearrow \qw & \ctrl{1} & \gate{\overset{\curvearrowright}{+}} & \curvearrowleft \qw & \ctrl{1} & \qw & \qw & \qw & \\
    & \curvearrowleft \qw & \gate{\sqrt{X}} & \nearrow \qw & \gate{\overset{\curvearrowleft}{-}} & \qw & \nearrow \qw & \gate{\overset{\curvearrowright}{+}} & \qw & \nwarrow \qw & \gate{\overset{\curvearrowright}{-}} & \qw & \nwarrow \qw & \gate{+} & \curvearrowright \qw & \gate{\sqrt{X}} & \circlearrowleft \qw\\
    }
\end{array}\]
\caption{Final optimized Toffoli gate for linearly connected qubit architectures.}
\label{fig:lc_optimal_final}
\end{figure}
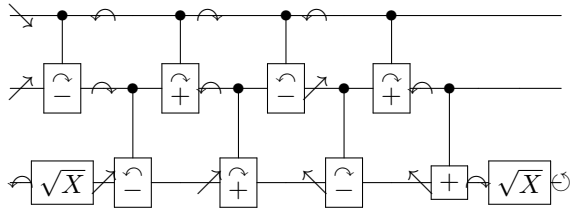

The application of native gate-level optimization resulted in the Toffoli gate implementation in \Cref{fig:lc_optimal_final}. Please refer to \Cref{fig:sandwiched} to interpret the quantum circuit diagram. \Cref{tab:lc_toffoli_improvements} compares the canonical and optimized Toffoli gate implementations using three fidelity proxies. Fidelity proxies are circuit properties that are correlated with average circuit fidelity. Generally, the fidelity of a quantum circuit improves as the circuit length, total rotation, and the number of native gates decrease. Circuit length is the time taken to execute the critical path of a quantum circuit; total rotation is the total degrees of x-rotations required to implement all single-qubit gates in a quantum circuit. Prior work suggests that circuit length is most correlated with average circuit fidelity in the NISQ era~\cite{cheng2020accqoc}. We note that $R_z$ rotations
are not counted in the number of native gates because they are not
physically executed operations.

\begin{table}[h]
\centering
\caption{\label{tab:lc_toffoli_improvements}Optimization proxy improvements of the optimized Toffoli gate for linearly connected qubit architectures.}
\begin{tabular}{llll}
\hline \hline
\textbf{Proxy} & \textbf{Canonical} & \textbf{Optimized} & \textbf{Change}\\
\hline
Length (ns) & $2694.1$ & $2303.0$ & $-14.5\%$\\
Total rotation & $3960^{\circ}$ & $2250^{\circ}$ & $-43.2\%$\\
Native gates & 28 & 17 & $-39.3\%$\\
\end{tabular}
\end{table}

\subsection{Benchmarking}
To demonstrate the efficacy of the optimized Toffoli gate implementation in \Cref{fig:lc_optimal_final}, we benchmarked it on qubits 0, 1, and 3 of the IBMQ  Jakarta quantum computer, using SuperstaQ \cite{superstaq} to produce the optimized Toffoli gate. While truth tables can verify the correctness of classical logic gates, benchmarking the performance of quantum gates presents novel challenges. Examining the truth table of a quantum gate over the computational basis
states, for example, does not encapsulate its performance on states that are
superpositions of these states. Superposition states represent a much larger
portion of the Hilbert space on which a quantum gate operates, making it
important to include them in any complete characterization of a quantum gate.
One prominent benchmarking method used to characterize the accuracy of a
quantum gate is quantum process tomography.

In quantum process tomography, many representative states are
prepared and measured to reconstruct the true operation being performed. In
Qiskit, each qubit of the operation to be benchmarked is prepared in one of the
four Pauli-basis eigenstates and then measured in the $X$, $Y$, or $Z$ Pauli
basis. This yields $4^n3^n=12^n$ total benchmarking circuits for an $n$-qubit
operation. Each of these circuits is then run multiple times on quantum
hardware, and the resulting outcome distributions are used to build a Choi
matrix representing the reconstructed action of the quantum channel. Qiskit
then calculates a real number in the range $[0, 1]$ termed average gate
fidelity using \Cref{eq:fidelity}, where $\mathcal{E}$ is the reconstructed Choi operator and $U$ is the target unitary~\cite{Qiskit}. The closer this value is to $1$, the better the quantum process truly matches its target unitary. The average gate fidelity offers a means of comparison between implementations of quantum gates. We note that quantum process tomography, unlike other methods such as randomized benchmarking, is not immune to state preparation and measurement errors. To reduce the effects of measurement errors, we ran calibration circuits to build a correction matrix for adjusting quantum hardware distributions.

\begin{equation}
    F_{avg}(\mathcal{E}, U)=\int d\psi \bra{\psi}U^\dag\mathcal{E}(\ket{\psi} \bra{\psi}) U \ket{\psi}
\label{eq:fidelity}
\end{equation}

In order to establish the statistical significance of our optimizations, error bounds on average fidelities were also computed. The error bounds in the gate fidelities are obtained by a Monte Carlo method~\cite{barends2014superconductingmontecarlo, huang2019fidelitymontecarlo2, noiri2022fastmontecarlo3} assuming that the measured probabilities follow multinomial distributions. We randomly sample the single-shot probabilities 1,000 times to obtain the gate fidelity distribution. The error bounds are defined as 1.96 times of standard deviation or the 95\% confidence level of a least-squares fit.

\begin{figure}
    \centering
    \includegraphics[width=0.45\textwidth]{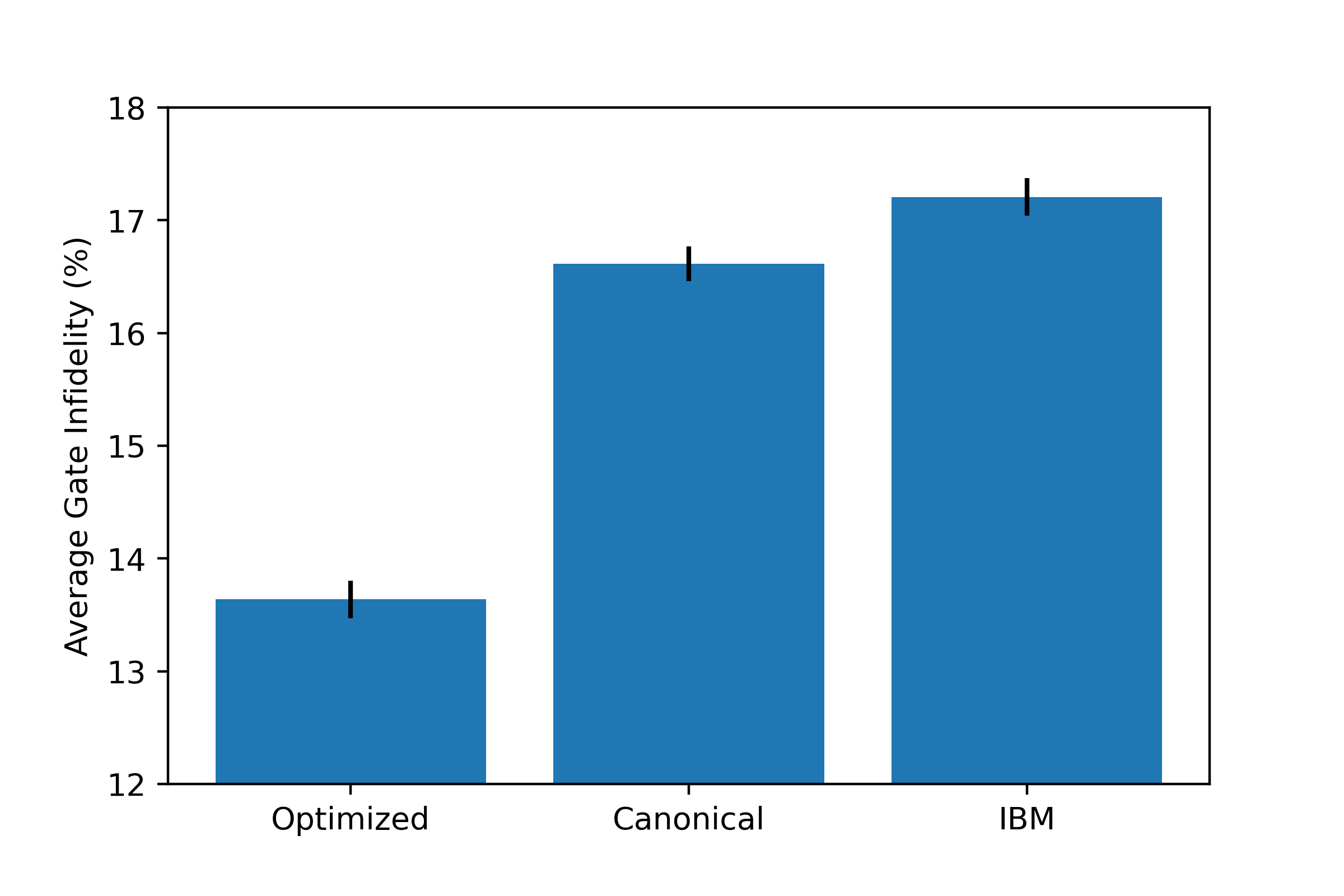}
    \caption{Quantum process tomography benchmarked infidelity of the native gate-level optimized Toffoli gate produced with SuperstaQ, compared with the infidelities of the canonical implementation and IBM-optimized implementation.}
    \label{fig:qpt}
\end{figure}

\Cref{fig:qpt} compares the average fidelity of our optimized Toffoli gate implementation with the average fidelities of the canonical linear Toffoli gate implementation (shown in \Cref{fig:lc_toffoli}) and IBM Qiskit-optimized linear Toffoli gate implementation (compiled with optimization level $3$ and transpilation seed $12345$). We measured our optimized Toffoli gate, the canonical Toffoli gate, and the IBM Qiskit-compiled Toffoli gate average gate fidelities to be $0.864 \pm 0.00166$, $0.834 \pm 0.00153$, and $0.828 \pm 0.00166$, respectively. Therefore, our optimization results in an $18\%$ reduction in infidelity compared with the canonical Toffoli gate.

\section{Numerical Discovery Approach}\label{sec:numerical}
The methods presented in \Cref{sec:apriori} focused on the refinement of an already existing quantum circuit. In this section, we will focus on the discovery of novel quantum circuit structures for implementing a target unitary matrix using numerical optimization techniques. After an exploration of our mathematical methodology and its results for the Toffoli gate, we will discuss how the numerical quantum circuit discovery approach can be integrated with the a priori analytical methods presented in the preceding section.

\subsection{Methods}\label{sec:numerical_methods}
We now turn our attention to developing numerical methods for quantum circuit discovery. First, we will present a technique for parameterizing the set of all quantum circuits containing $n$ multi-qubit gates. Then, we will present an objective function leveraging this parameterization which, when minimized, produces a quantum circuit implementing a specified unitary. Finally, we will describe how to use this objective function to computationally find quantum circuit implementations of unitary matrices in a parallel, massively scalable fashion. Although prior work has explored numerical optimization for quantum circuit synthesis~\cite{resynthesis2021}, the novelty of this work lies in the objective function used. The objective function presented in this work allows for the combination of numerical circuit synthesis with a priori analytical methods.

\subsubsection{Parameterizing Quantum Circuits}\label{sec:param}
We first turn our attention to establishing a method of parameterizing the set of all quantum circuits. Note that any sequence of single-qubit operations can be expressed as the product of several unitary matrices, which itself is a unitary matrix. Any single-qubit operation can be expressed via a generic single-qubit rotation gate, denoted by $U$. The unitary matrix of the $U$ gate can be completely specified with three real numbers $\theta$, $\phi$, and $\lambda$~\cite{Nielsen2009}. The $U$ gate is realized at the native gate level with the sequence $R_z(\phi)R_z(180^\circ)\sqrt{X}R_z(180^\circ)R_z(\theta)\sqrt{X}R_z(\lambda)$ on IBMQ hardware.

It follows that sequences of arbitrary single-qubit gates occurring between
multi-qubit gates can be replaced by one layer of $U$ gates. Rearranging the
quantum circuit in this manner results in $n+1$ layers of single-qubit gates
interleaved with $n$ multi-qubit operations. Now, group every multi-qubit
operation with the corresponding preceding layer of $U$ gates, and call this a
block. Therefore, any circuit containing $n$ multi-qubit gates can be expressed
by using $n$ blocks followed by a layer of $U$ gates. We use $B_0$ to denote
the unitary matrix representing the first layer of $U$ gates. The blocks that can be used as building blocks are determined by the native gate set over which one
optimizes. The unitary describing the action of the entire circuit equals $B_n
B_{n-1} \dots B_0$ (note that multiplied unitary matrices are applied to a
quantum state from right to left).

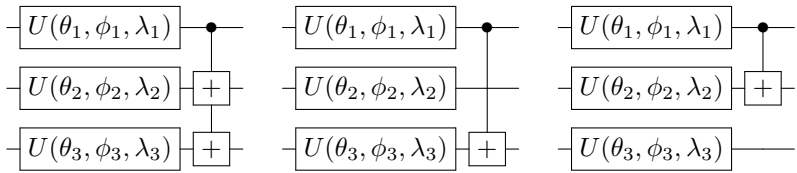
\begin{figure}[h]
\[\begin{array}{l}
    \Qcircuit @C=0.5em @R=.7em {
    & & \gate{U(\theta_1, \phi_1, \lambda_1)} & \ctrl{1} & \qw \\
    & & \gate{U(\theta_2, \phi_2, \lambda_2)} & \gate{+} \qwx[1] &  \qw\\
    & & \gate{U(\theta_3, \phi_3, \lambda_3)} & \gate{+} & \qw \\
    }
    \Qcircuit @C=0.5em @R=.7em {
    & & & & \gate{U(\theta_1, \phi_1, \lambda_1)} & \ctrl{2} & \qw\\
    & & & & \gate{U(\theta_2, \phi_2, \lambda_2)} & \qw & \qw\\
    & & & & \gate{U(\theta_3, \phi_3, \lambda_3)} & \gate{+} & \qw\\
    }
    \Qcircuit @C=0.5em @R=.7em {
    & & & & \gate{U(\theta_1, \phi_1, \lambda_1)} & \ctrl{1} & \qw\\
    & & & & \gate{U(\theta_2, \phi_2, \lambda_2)} & \gate{+} & \qw\\
    & & & & \gate{U(\theta_3, \phi_3, \lambda_3)} & \qw & \qw \\
    }
\end{array} \]
\caption{The three possible layers implementable on a linear three-qubit architecture with qubit $q_0$ connected to qubits $q_1$ and $q_2$.}
\label{fig:layers}
\end{figure}

Optimizing with linear qubit connectivity poses constraints on which block
unitaries are allowed, reducing the optimization space. Note that multi-target cross-resonance (MCR) gates{} 
must all have the same control bit, for example, because allowing MCR gates
with different control bits would require a fully connected qubit architecture.
Let us label a set of three linearly connected qubits $q_0$, $q_1$, and $q_2$.
We arbitrarily choose qubit $q_0$ to be the MCR control bit
without loss of generality. This reduces our gate set to single-qubit gates,
the MCR gate with $q_0$ as the control qubit, the CR gate between
$q_0$ and $q_1$, and the CR gate between $q_0$ and $q_2$, because a CR
gate between any other qubit pairs would violate the imposed connectivity
constraint. \Cref{fig:layers} illustrates the three blocks that can 
construct all possible quantum circuits containing $n$ multi-qubit
operations over this gate set. This rigorous, parameterized approach to
describing quantum circuits allows us to formulate our principal goal in the
language of numerical optimization.

\subsubsection{Objective Formulation}
We can frame the problem of Toffoli optimization over a linear qubit
architecture in the language of unconstrained nonlinear optimization. Consider
an arbitrary $n$-layer quantum circuit parameterized via the method presented
in \Cref{sec:param}. Let the $U$ gate parameters in this circuit be organized into a vector in $\mathbb{R}^{3n+3}$ as shown in \Cref{eq:beta}.

\begin{equation}
    \vec{\beta} =
\rvect{\theta_1 & \phi_1 & \lambda_1 & \cdots & \theta_{3n+3} & \phi_{3n+3} &
\lambda_{3n+3}}^{T}
\label{eq:beta}
\end{equation}

Let $U(\vec{\beta})=B_n B_{n-1} B_0$ represent the final unitary matrix obtained from multiplying parameterized block unitary matrices as a function of $\vec{\beta}$. We wish to solve the nonlinear least-squares problem detailed in \Cref{eq:nlsprob}.

\begin{equation}
    \vec{\beta^*}=\operatorname*{argmin}_{\vec{\beta}} \| U_{CCX} - U(\vec{\beta}) \|_F^2
\label{eq:nlsprob}
\end{equation}

\begin{figure}
    \centering
    \includegraphics[width=0.5\textwidth]{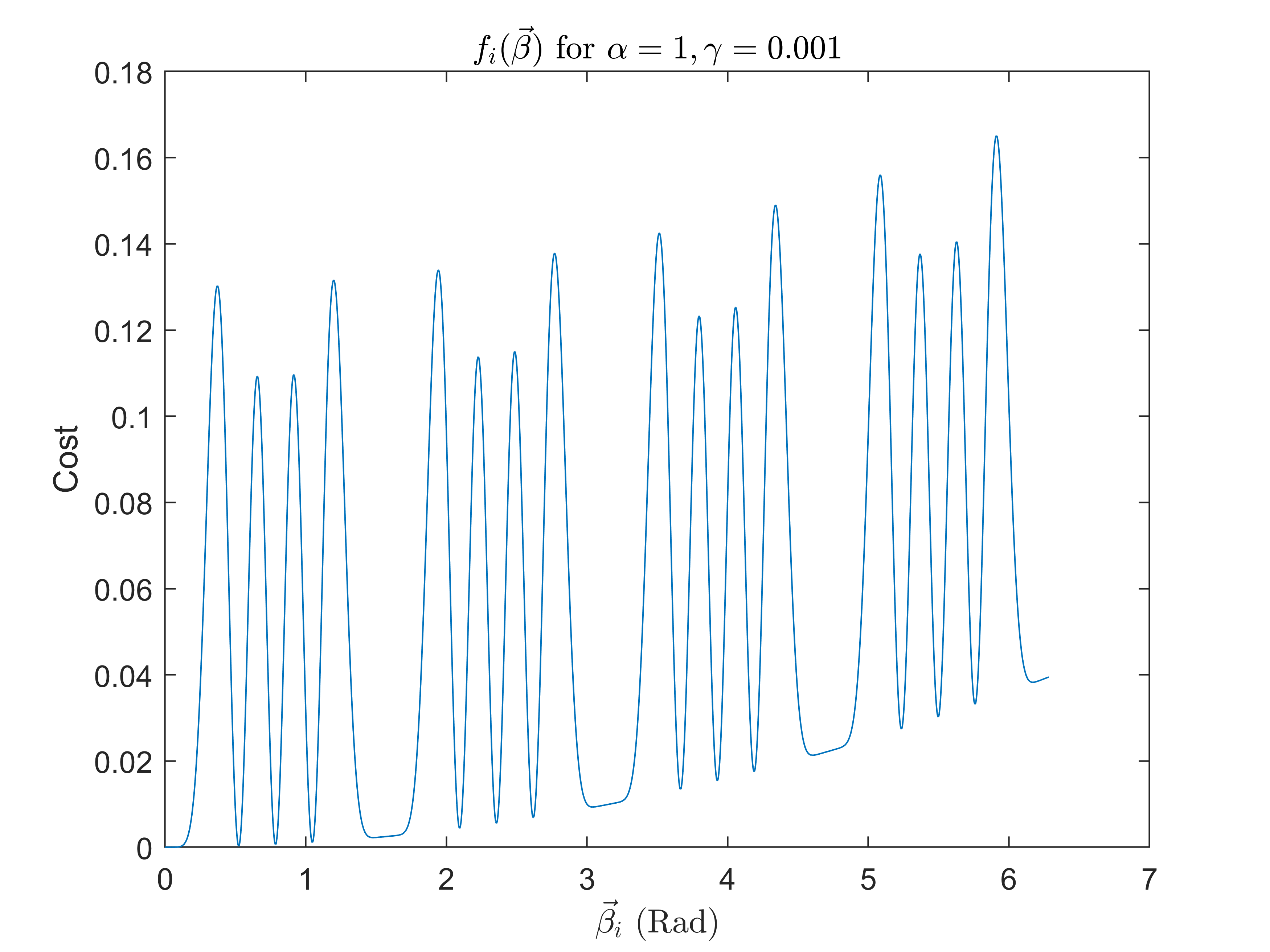}
    \caption{$f_i(\vec{\beta})$ with parameters $\alpha=1$ and $\gamma=0.001$.}
    \label{fig:beta_cost_picture}
\end{figure}

Note that this \Cref{eq:nlsprob}, by definition of the
Frobenius norm, is the sum of squared residuals and is a measure of how close $U(\vec{\beta})$ is to implementing the Toffoli gate. While this objective alone is sufficient for Toffoli circuit discovery, it has limited applicability in that the parameter values produced by optimizing it can take on any real angle value. This makes it nearly impossible to apply analytical a priori methods (such as cross-gate pulse cancellation and active rotation virtualization) to numerically discovered Toffoli circuits. In order to make these techniques complementary, a more complex objective function penalizing certain parameter values and favoring others is required. \Cref{eq:beta_cost} presents a function that favors small parameter angle values that are multiples of $\frac{\pi}{6}$, $\frac{\pi}{4}$, and $\frac{\pi}{3}$. \Cref{fig:beta_cost_picture} graphically depicts $f_i(\vec{\beta})$ with parameters $\alpha=1$ and $\gamma=0.001$.

\begin{equation}
     f(\vec{\beta})= \alpha \sum_{i=1}^{9n+9} \sin^2(6 \vec{\beta}_i) \sin^2(4 \vec{\beta}_i) \sin^4(2 \vec{\beta}_i) + \gamma \|\vec{\beta} \|^2
     \label{eq:beta_cost}
\end{equation}

Combining \Cref{eq:nlsprob} and \Cref{eq:beta_cost} results in the objective function in \Cref{eq:objective_final}. Tuning parameters $\alpha$ and $\gamma$ scale the importance of small, ideal angles relative to accuracy.

\begin{equation}
    \vec{\beta^*}=\operatorname*{argmin}_{\vec{\beta}} \| U_{CCX} - U(\vec{\beta}) \|_F^2 + f(\vec{\beta})
    \label{eq:objective_final}
\end{equation}

This problem must be solved for every permutation of block types. Because there are three block choices per layer for a system with linear qubit connectivity, $3^n$ nonlinear least-squares problems must be solved to adequately search for Toffoli gate implementations with $n$ multi-qubit operations. Since the shortest known implementation of the Toffoli gate over the gate set including single-qubit operations and the CR gate consists of eight CR gates, we need only search $n=1,2,3,\dots,7$, since $n \ge 8$ would not generate more optimal linear Toffoli gate implementations.

For this numerical optimization problem, one must consider how close the unitary generated by a set of optimal parameters must be to the true unitary in order to be considered a Toffoli gate implementation. It is generally accepted that if the fidelity of a unitary with respect to some target unitary is above 99.9\% (or, equivalently, the infidelity of a unitary with respect to some target unitary is below 0.1\%), then the unitary implements the target unitary~\cite{Gokhale_2019}. Infidelity in this sense~\cite{Leung_2017} is defined in \Cref{eq:infidelity} where $\mu_{1}$ is infidelity, $U_{CCX}$ is the Toffoli gate unitary matrix, and $U(\vec{\beta})$ is the unitary matrix of a particular Toffoli gate implementation.

\begin{equation}
    \mu_{1} = 1 - \left|  \frac{\Tr(U_{CCX}^\dag U(\vec{\beta}))}{8} \right| ^2
    \label{eq:infidelity}
\end{equation}

\subsubsection{Minimizing the Objective Function}
To find parameters that minimize the objective function in \Cref{eq:objective_final}, we used the Levenberg-Marquardt algorithm~\cite{more1978levenberg}. This algorithm was chosen because it is a standard method of solving unconstrained nonlinear optimization problems with several readily available software implementations. The Levenberg-Marquardt algorithm is a gradient-based algorithm that requires an initial guess from which to descend. Selection of the initial guess results in varying success, so several initial points per layered configuration must be tested. In practice, it was found that on the order of tens of points was sufficient.

\begin{algorithm2e}
\KwIn{list $o$ of multi-qubit native gates available on target hardware, integer $n \ge 1$ specifying the number of multi-qubit native gates that implementations should contain, and integer $k \ge 1$ specifying the number of times to check each circuit configuration}
\KwOut{$\vec{\beta}^* \in \mathbb{R}^{9n+9}$}

$\vec{\beta}^* \leftarrow $ null 

\ForEach{$n$-length quantum circuit structure constructed from $o$} {
    \For{$i\leftarrow 1$ \KwTo $k$} {
        $\vec{\beta}_0 \longleftarrow \vec{v} \in \mathbb{R}^{9n+9}$ where each entry in $\vec{v}$ is chosen from a uniform random distribution over the domain $[0, 2\pi)$ 
        
        $\vec{\beta} \leftarrow$ result of running Levenberg-Marquardt starting from $\vec{\beta}_0$ 
        
        \If{$\vec{\beta}^* =$ {\rm null} $\lor$ $(U(\vec{\beta}) \|_F^2 + f(\vec{\beta}) \le U(\vec{\beta}) \|_F^2 \land f(\vec{\beta}) \land U(\vec{\beta}) \|_F^2 \leq 0.01)$} {
            $\vec{\beta}^* \leftarrow \vec{\beta}$ 
        }
    }
}

\Return $\vec{\beta}^*$ 
\caption{Searching for $n$-layer Toffoli implementation for a native gate set.}
\end{algorithm2e}

The computational complexity of this algorithm is $O(n^2 3^n)$ where $n$ is the number of circuit layers. Although this algorithm is exponential in nature, it is important to note the ease with which it can be parallelized. For example, a cluster of machines could each be assigned a specific multi-qubit circuit configuration to explore. If the machines have multiple cores, each core can be assigned a different initial parameter guess. The runtime of the ideal critical path length of this computation scales at rate $O(n^2)$.  Although this algorithm can be computationally expensive, once novel implementations are discovered (e.g., via considerable computational effort), they can be easily stored and used by compilers on personal computers.

Additionally, if the desired underlying multi-qubit configuration of a circuit is already known, the algorithm can be run sequentially in polynomial time. Therefore, an additional application of this algorithm for larger circuits is reparameterizing them to enable analytical a priori optimizations.

\subsection{Results}
The algorithm presented in \Cref{sec:numerical_methods} was implemented in C++~\cite{QuOpt} using the Google Ceres library~\cite{ceres-solver} and parallelized with MPI. It was then used to run a computational search
of $n$-layer Toffoli gate implementations for $n=1,2,3,\ldots,7$ on a personal workstation (for $n=1,2,3,4$) and on the Argonne Bebop supercomputer (for $n=5,6,7$). On the supercomputer, the program ran on $10$ nodes. Each node solved all $3^n$ optimization problems on $36$ cores using the Levenberg--Marquardt algorithm and a randomly generated starting point. For $n=6$, the distributed algorithm ran for roughly half an hour. This implementation used an objective function agnostic to angle parameters (equivalent to setting $\alpha=\gamma=0$ in \Cref{eq:objective_final}) and discovered $38$ six-layer quantum circuit structures and corresponding parameters capable of implementing the Toffoli gate.

\begin{figure}
    \[\begin{array}{l}
     \Qcircuit @C=1em @R=.5em {
     & \gate{U} & \ctrl{2} & \gate{U} & \ctrl{1} & \gate{U} & \ctrl{1} & \gate{U} & \ctrl{1} & \gate{U} & \ctrl{1} & \gate{U} & \ctrl{1} & \gate{U} & \qw \\
     & \gate{U} & \qw & \qw & \gate{+} & \gate{U} & \gate{+} \qwx[1] & \gate{U} & \gate{+} \qwx[1] & \gate{U} & \gate{+} \qwx[1] & \gate{U} & \gate{+} & \gate{U} & \qw &\\
     & \gate{U} & \gate{+} & \gate{U} & \qw & \qw & \gate{+} & \gate{U} & \gate{+} & \gate{U} & \gate{+} & \gate{U} & \qw & \qw & \qw &\\
     }
     \end{array} \]
 \caption{Multi-qubit gate structure of Toffoli gate implementation for linearly connected qubits.}
 \label{fig:bebop_candidate}
\end{figure}
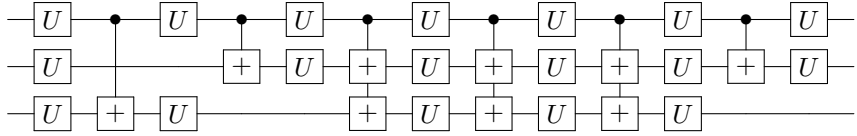

One of these quantum circuit structures, shown in \Cref{fig:bebop_candidate} with angle parameters omitted for clarity, was then reparameterized by applying the SciPy Levenberg-Marquardt least squares optimizer~\cite{2020SciPy-NMeth} to the objective function in \Cref{eq:objective_final} with $\alpha=0.30$ and $\gamma=10^{-8}$. The starting parameter vector provided to the least squares optimizer greatly affects the solution to which it will converge. By selecting several random starting parameter vectors and running the optimizer, it was found that this quantum circuit structure yielded several valid parameterizations with desirable angles (code and angle parameter values are available on GitHub~\cite{PyQuOpt}). Therefore, we propose numerical reparameterization of quantum circuit subsections as a tool for producing optimization opportunities. %

\subsection{Connection to Analytical Approach} %
\begin{figure}[h]
\[\begin{array}{l}
    \Qcircuit @C=1em @R=.7em {
    & q_o & & \gate{H} & \ctrl{1} & \qw & \gate{H} & \qw \\
    & q_1 & & \qw & \targ &  \qw & \gate{X} & \qw \\
    }       
    \Qcircuit @C=1em @R=.7em {
    & & & & q_o & & \gate{U(\pi, \frac{5\pi}{4}, \frac{3\pi}{4})} & \ctrl{1} & \qw & \qw \\
    & & & & q_1 & & \gate{U(\frac{\pi}{2}, \frac{3\pi}{2}, \frac{\pi}{2})} & \gate{+} &  \qw & \qw \\
    }
\end{array} \]
\caption{Simple quantum circuit (left) and its optimized implementation (right) found by using both numerical discovery and analytical a priori methods.}
\label{fig:opt_example}
\end{figure}
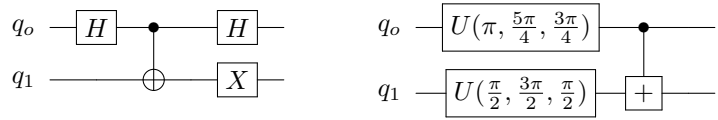

It should be underscored that the numerical discovery approach and analytical a priori approach are complementary. The numerical discovery approach can be used to find an initial implementation of a circuit or subcircuit, which can then be refined using analytical a priori methods. This is best illustrated with an example. Consider the simple two-qubit circuit in \Cref{fig:opt_example}.

Notice that upon transpilation to native gates, there is no clearly applicable analytical a priori method. Cross-gate pulse cancellation, for example, fails because $R_z$ gates used to implement the Hadamard and CNOT gates block physical x-rotation gates from canceling. It is in this situation that the numerical discovery method should be used. Notice this circuit is a two-qubit GHZ quantum circuit followed by a Hadamard gate and an $X$ gate. If there exists a quantum circuit producing the two-qubit GHZ state that ends with a native gate sequence equivalent to a Hadamard gate on $q_0$ and an $X$ gate on $q_1$, then cross-gate pulse cancellation may be applied to entirely cancel the final two single-qubit gates in this circuit since both gates are their own inverses.

Such an implementation does in fact exist. It was discovered numerically using the methods developed in this section and led to the optimized circuit found in \Cref{fig:opt_example}. Note that this implementation requires only two layers of $\sqrt{X}$ gates and $540^\circ$ of active rotation. This compares favorably to the original quantum circuit, which contained five layers of $\sqrt{X}$ gates and $990^\circ$ degrees of active rotation.

This example clearly demonstrates the complementary nature of analytical a priori techniques such as cross-gate pulse cancellation and numerical quantum circuit discovery. The numerical discovery approach can be integrated into existing quantum compilers to provide otherwise unavailable optimization opportunities.

While the numerical discovery method is best suited for optimizing subcircuits with a small number of qubits, it can be easily generalized to an arbitrary number of qubits and multi-qubit gate set by determining the appropriate fundamental "layers" (see \Cref{fig:layers}). The applicability of these techniques to any native gate set and superconducting qubit architecture makes the numerical discovery approach broadly applicable to the quantum computing community.

\section{Application to Quantum Compilation}\label{sec:applications}

Quantum compilation is a highly non-trivial task that involves solving several problems, including qubit routing and decomposition to native gates. Traditionally, quantum compilation follows a two-pass approach in which input quantum circuits are first decomposed and then mapped onto physical qubits. Recently, however, a three-pass approach called "Orchestrated Trios" has been proposed~\cite{Duckering2021}. In this approach, the compiler decomposes a quantum circuit while leaving Toffoli gates intact in its first pass. Next, the qubits acted on by Toffoli gates are routed (i.e., moved) in groups, or "trios" to accommodate the qubit architecture of the target device. Finally, all Toffoli gates are decomposed into either the standard 6-CNOT or 8-CNOT implementations, depending on trio connectivity.

The Toffoli gate optimization presented in \Cref{sec:apriori} of this paper was integrated into the third pass of the Orchestrated Trios compiler (referred to as the "PulseTrios" compiler). Toffoli gates were decomposed into the native gate-level quantum circuit in \Cref{fig:lc_optimal_final}. This resulted in significant quantum circuit length reductions. 
\Cref{tab:orchestrated_trios} shows the estimated probability of success of several quantum circuit benchmarks~\cite{BakerBenchmarks} and QAOA circuits for Maximum Independent Set~\cite{tomesh2022qaoa2} compiled with Qiskit, Trios, and PulseTrios.

The majority of these benchmarks have large depths and very low fidelities on current quantum hardware, making quantum process tomography infeasible to perform. Therefore, scaling system parameters such as gate error and decoherence time constants is a standard approach for estimating the probability of success of quantum circuits on near-term quantum devices. To illustrate the results of the optimized Toffoli gates on benchmark success rates, the IBMQ Mumbai device was chosen as the target device. We assume gate error rates to be $10$x smaller than presently measured and decoherence time constants to be $10$x larger than presently measured. Under these assumptions, the probability of quantum circuit success on a particular system can be estimated with \Cref{eq:algorithm_fidelity}, where $F_i$ is the average fidelity of gates of type $i$, $n_i$
is the number of gates of type $i$ used, $\Delta$ is total circuit run time, $T_1$ is the system thermal relaxation time constant, and $T_2$ is the system dephasing time constant.

 While PulseTrios shows an application of our purely analytical optimizations to quantum compilation, our combined analytical and numerical approach has been demonstrated by the QContext quantum compiler\cite{QContext} on the IBMQ native gate set. QContext used the numerical discovery software presented in \Cref{sec:numerical} to synthesize new implementations of the CNOT gate that yield cross-gate pulse cancellation opportunities. These cancellation opportunities resulted in QContext outperforming OrchestratedTrios, the state-of-the-art quantum compiler.

\begin{equation}
    P_{success} = e^{-10\Delta/T_1 - 10\Delta/T_2} \prod_{i=1}^{N} \left(1 - \frac{1 - F_i}{10} \right) ^{n_i}
\label{eq:algorithm_fidelity}
\end{equation}

\begin{table}
\caption{\label{tab:orchestrated_trios}Projected success probabilities of several quantum circuit benchmarks~\cite{BakerBenchmarks} when compiled with Qiskit, Trios, and PulseTrios on near-term hardware. The geometric mean of the improvement of PulseTrios over Qiskit and Trios is shown in the final column.}
\centering
\begin{tabular}{l|cccc|ccc|c}
\hline \hline
\textbf{Benchmark} & \textbf{\#Qubits} &\textbf{\#Toffoli} &\textbf{\#CNOT}  &\textbf{\#U3} & \textbf{Qiskit} & \textbf{Trios} & \textbf{PulseTrios} & \textbf{Improvement}\\
\hline
cnx\_dirty~\cite{baker2019decomposing} & 18 & 16 & 0& 0& 66.0\% & 74.4\% & 78.1\% & 1.09x\\
cnx\_halfborrowed~\cite{gidney_2015} & 19 & 32 & 0 & 0& 47.2\% & 56.9\% & 62.7\% & 1.15x\\
cnx\_logancilla~\cite{Barenco_1995} & 19 & 17 & 0 & 0 & 64.4\% & 73.1\% & 77.0\% & 1.09x\\
cnx\_inplace~\cite{gidney_2015} & 4 & 54 & 46 & 33 & 29.6\% & 36.2\% & 42.6\% & 1.20x\\
cuccaro\_adder~\cite{quant-ph/0410184} & 20 & 18 & 46 & 18 & 55.7\% & 67.7\% & 71.5\% & 1.13x\\
takahashi\_adder~\cite{takahashi2009quantum} & 20 & 18 & 44 & 0 & 56.0\% & 68.9\% & 72.7\% & 1.14x\\
borrowedbit~\cite{gidney_2015} & 5 & 50 & 38 & 21 & 32.4\% & 39.4\% & 45.8\% & 1.19x\\
grovers~\cite{Grover1996} & 9 & 84 & 0 & 78 & 8.9\% & 22.3\% & 28.9\% & 1.80x\\
QAOA\_node6~\cite{hadfield2019qaoa1, tomesh2022qaoa2} & 6 & 4 & 32 & 56 & 81.5\% & 87.8\% & 88.9\% & 1.04x\\
QAOA\_node8~\cite{hadfield2019qaoa1, tomesh2022qaoa2} & 8 & 8 & 148 & 195 & 52.8\% & 73.3\% & 75.1\% & 1.19x\\
\end{tabular}
\end{table}

\section{Conclusion}\label{sec:conclusion}
Native gate-level analysis has opened a vast array of opportunities for optimization. In this paper we demonstrated that examining
Toffoli gate implementations in the native gate set of target hardware (in this case IBMQ machines) yields several opportunities for quantum circuit length reduction. This paper presented cross-resonance gate sandwiching as a new analytical optimization technique that significantly reduces quantum circuit length. This technique was used in conjunction with existing techniques to produce a linear Toffoli gate implementation with $18\%$ lower infidelity.

This paper also contributed methodology for parameterizing quantum circuits with a finite set of real numbers and formulating quantum circuit length reduction as a nonlinear least-squares optimization problem that can be solved numerically in a distributed fashion. This technique was used to discover a Toffoli gate implementation with two fewer required multi-qubit gates than standard on a linearly connected qubit architecture with the IBMQ native gate set augmented with the multi-target cross-resonance gate. While this implementation was not able to be benchmarked with quantum process tomography, it merits further exploration and represents a direction for future research. In order to facilitate further exploration, the optimization code developed for this paper will be made publicly available on GitHub.

Although optimizations were performed for the IBM system because (a) it is on of the most widely used
quantum hardware platform and (b) it is a leading system in terms of
low-level access to pulse control and native gates, the principles
expressed in our manuscript are applicable to other hardware platforms. For
instance, the hidden inverse technique\cite{zhang2022hidden} has been used on
trapped-ion systems; this technique is similar to the echo in 
cross-resonance gates and can be similarly optimized. In the same spirit, the
leading two-qubit gate protocol on neutral atom systems\cite{levine2019parallel}
also involves an echoed sequence, which creates a polarity freedom that can be
optimized.

Several avenues of future exploration exist for this topic. One direction is the continued development of context-aware compilers that leverage numerical resynthesis. Another direction is the application of the methodology presented in this paper to quantum gates and circuits besides the Toffoli gate. In terms of the native gate set augmentation aspect of this paper, the multi-target cross-resonance gate represents only one of several promising gates that could augment the IBMQ native gate set
architecture; several other
native gate set augmentation candidates can be explored.

In summary, our work on optimizing the Toffoli gate supports the idea that examining quantum gates at a low level of abstraction yields a surprising
number of opportunities for performance improvements. The primary contribution
of this paper is to encourage further exploration of quantum circuit
optimization at the native gate set level of abstraction by concretely
demonstrating the improvement of a frequently used quantum gate.

\section*{Acknowledgement}
This material is based upon work supported by the U.S. Department of Energy, Office of Science, under contract number DE-AC02-06CH11357. The work of M.B.~was supported in part by the U.S.~Department of Energy, Office of Science, Office of Workforce Development for Teachers and Scientists (WDTS) under the Science Undergraduate Laboratory Internship (SULI) program. The work of M.B., M.S., and J.L.~was supported by Q-NEXT, one of the U.S.~Department of Energy Office of Science (DOE-SC) National Quantum Information Science Research Centers and the Office of Advanced Scientific Computing Research, Accelerated Research for Quantum Computing program. P.G. acknowledges US Department of Energy funding from the Advanced Manufacturing Office (CRADA No.~2020-20099.) and the Office of Science, Office of Advanced Scientific Computing Research (Award Number DE-SC0021526). This research used resources of the Oak Ridge Leadership Computing Facility, which is a DOE Office of Science User Facility supported under Contract DE-AC05-00OR22725. We also acknowledge the computing resources provided on Bebop, a high-performance computing cluster operated by the Laboratory Computing Resource Center at Argonne National Laboratory.

\bibliographystyle{unsrtnat}
\bibliography{references}

\begin{thebibliography}{61}
\providecommand{\natexlab}[1]{#1}
\providecommand{\url}[1]{\texttt{#1}}
\expandafter\ifx\csname urlstyle\endcsname\relax
  \providecommand{\doi}[1]{doi: #1}\else
  \providecommand{\doi}{doi: \begingroup \urlstyle{rm}\Url}\fi

\bibitem[Preskill(2021)]{preskill2021quantum}
John Preskill.
\newblock Quantum computing 40 years later.
\newblock Technical Report 2106.10522, arXiv, 2021.
\newblock URL \url{https://arxiv.org/abs/2106.10522}.

\bibitem[Arute et~al.(2019)Arute, Arya, Babbush, Bacon, Bardin, Barends,
  Biswas, Boixo, Brandao, Buell, Burkett, Chen, Chen, Chiaro, Collins,
  Courtney, Dunsworth, Farhi, Foxen, Fowler, Gidney, Giustina, Graff, Guerin,
  Habegger, Harrigan, Hartmann, Ho, Hoffmann, Huang, Humble, Isakov, Jeffrey,
  Jiang, Kafri, Kechedzhi, Kelly, Klimov, Knysh, Korotkov, Kostritsa, Landhuis,
  Lindmark, Lucero, Lyakh, Mandr{\`{a}}, McClean, McEwen, Megrant, Mi,
  Michielsen, Mohseni, Mutus, Naaman, Neeley, Neill, Niu, Ostby, Petukhov,
  Platt, Quintana, Rieffel, Roushan, Rubin, Sank, Satzinger, Smelyanskiy, Sung,
  Trevithick, Vainsencher, Villalonga, White, Yao, Yeh, Zalcman, Neven, and
  Martinis]{Arute2019}
Frank Arute, Kunal Arya, Ryan Babbush, Dave Bacon, Joseph~C. Bardin, Rami
  Barends, Rupak Biswas, Sergio Boixo, Fernando G. S.~L. Brandao, David~A.
  Buell, Brian Burkett, Yu~Chen, Zijun Chen, Ben Chiaro, Roberto Collins,
  William Courtney, Andrew Dunsworth, Edward Farhi, Brooks Foxen, Austin
  Fowler, Craig Gidney, Marissa Giustina, Rob Graff, Keith Guerin, Steve
  Habegger, Matthew~P. Harrigan, Michael~J. Hartmann, Alan Ho, Markus Hoffmann,
  Trent Huang, Travis~S. Humble, Sergei~V. Isakov, Evan Jeffrey, Zhang Jiang,
  Dvir Kafri, Kostyantyn Kechedzhi, Julian Kelly, Paul~V. Klimov, Sergey Knysh,
  Alexander Korotkov, Fedor Kostritsa, David Landhuis, Mike Lindmark, Erik
  Lucero, Dmitry Lyakh, Salvatore Mandr{\`{a}}, Jarrod~R. McClean, Matthew
  McEwen, Anthony Megrant, Xiao Mi, Kristel Michielsen, Masoud Mohseni, Josh
  Mutus, Ofer Naaman, Matthew Neeley, Charles Neill, Murphy~Yuezhen Niu, Eric
  Ostby, Andre Petukhov, John~C. Platt, Chris Quintana, Eleanor~G. Rieffel,
  Pedram Roushan, Nicholas~C. Rubin, Daniel Sank, Kevin~J. Satzinger, Vadim
  Smelyanskiy, Kevin~J. Sung, Matthew~D. Trevithick, Amit Vainsencher, Benjamin
  Villalonga, Theodore White, Z.~Jamie Yao, Ping Yeh, Adam Zalcman, Hartmut
  Neven, and John~M. Martinis.
\newblock Quantum supremacy using a programmable superconducting processor.
\newblock \emph{Nature}, 574\penalty0 (7779):\penalty0 505--510, 2019.
\newblock \doi{10.1038/s41586-019-1666-5}.
\newblock URL \url{https://doi.org/10.1038/s41586-019-1666-5}.

\bibitem[Wright et~al.(2019)Wright, Beck, Debnath, Amini, Nam, Grzesiak, Chen,
  Pisenti, Chmielewski, Collins, Hudek, Mizrahi, Wong-Campos, Allen, Apisdorf,
  Solomon, Williams, Ducore, Blinov, Kreikemeier, Chaplin, Keesan, Monroe, and
  Kim]{wright2019benchmarking}
K.~Wright, K.~M. Beck, S.~Debnath, J.~M. Amini, Y.~Nam, N.~Grzesiak, J.-S.
  Chen, N.~C. Pisenti, M.~Chmielewski, C.~Collins, K.~M. Hudek, J.~Mizrahi,
  J.~D. Wong-Campos, S.~Allen, J.~Apisdorf, P.~Solomon, M.~Williams, A.~M.
  Ducore, A.~Blinov, S.~M. Kreikemeier, V.~Chaplin, M.~Keesan, C.~Monroe, and
  J.~Kim.
\newblock Benchmarking an 11-qubit quantum computer.
\newblock \emph{Nature Communications}, 10\penalty0 (1):\penalty0 1--6, 2019.
\newblock URL \url{https://doi.org/10.1038/s41467-019-13534-2}.

\bibitem[Jurcevic et~al.(2021)Jurcevic, Javadi-Abhari, Bishop, Lauer, Bogorin,
  Brink, Capelluto, Günlük, Itoko, Kanazawa, Kandala, Keefe, Krsulich,
  Landers, Lewandowski, McClure, Nannicini, Narasgond, Nayfeh, Pritchett,
  Rothwell, Srinivasan, Sundaresan, Wang, Wei, Wood, Yau, Zhang, Dial, Chow,
  and Gambetta]{jurcevic2021demonstration}
Petar Jurcevic, Ali Javadi-Abhari, Lev~S Bishop, Isaac Lauer, Daniela~F
  Bogorin, Markus Brink, Lauren Capelluto, Oktay Günlük, Toshinari Itoko,
  Naoki Kanazawa, Abhinav Kandala, George~A Keefe, Kevin Krsulich, William
  Landers, Eric~P Lewandowski, Douglas~T McClure, Giacomo Nannicini, Adinath
  Narasgond, Hasan~M Nayfeh, Emily Pritchett, Mary~Beth Rothwell, Srikanth
  Srinivasan, Neereja Sundaresan, Cindy Wang, Ken~X Wei, Christopher~J Wood,
  Jeng-Bang Yau, Eric~J Zhang, Oliver~E Dial, Jerry~M Chow, and Jay~M Gambetta.
\newblock Demonstration of quantum volume 64 on a superconducting quantum
  computing system.
\newblock \emph{Quantum Science and Technology}, 6\penalty0 (2):\penalty0
  025020, 2021.
\newblock URL \url{https://doi.org/10.1088/2058-9565/abe519}.

\bibitem[ANIS et~al.(2021)ANIS, Abraham, AduOffei, Agarwal, Agliardi, Aharoni,
  Akhalwaya, Aleksandrowicz, Alexander, Amy, Anagolum, Arbel, Asfaw, Athalye,
  Avkhadiev, Azaustre, Banerjee, Banerjee, Bang, Bansal, Barkoutsos, Barnawal,
  Barron, Barron, Bello, Ben-Haim, Bevenius, Bhatnagar, Bhobe, Bianchini,
  Bishop, Blank, Bolos, Bopardikar, Bosch, Brandhofer, Brandon, Bravyi, Bronn,
  Bryce-Fuller, Bucher, Burov, Cabrera, Calpin, Capelluto, Carballo, Carrascal,
  Carriker, Carvalho, Chen, Chen, Chen, Chen, Chen, Chevallier, Cholarajan,
  Chow, Churchill, Claus, Clauss, Clothier, Cocking, Cocuzzo, Connor, Correa,
  Cross, Cross, Cross, Cruz-Benito, Culver, C{\'o}rcoles-Gonzales, D, Dague,
  Dandachi, Dangwal, Daniel, Daniels, Dartiailh, Davila, Debouni, Dekusar,
  Deshmukh, Deshpande, Ding, Doi, Dow, Drechsler, Dumitrescu, Dumon, Duran,
  EL-Safty, Eastman, Eberle, Ebrahimi, Eendebak, Egger, Espiricueta, Everitt,
  Facoetti, Farida, Fern{\'a}ndez, Ferracin, Ferrari, Ferrera, Fouilland,
  Frisch, Fuhrer, Fuller, GEORGE, Gacon, Gago, Gambella, Gambetta, Gammanpila,
  Garcia, Garg, Garion, Gates, Gil, Gilliam, Giridharan, Gomez-Mosquera,
  Gonzalo, de~la Puente~Gonz{\'a}lez, Gorzinski, Gould, Greenberg, Grinko,
  Guan, Gunnels, Gupta, G{\"u}nther, Haglund, Haide, Hamamura, Hamido, Harkins,
  Hasan, Havlicek, Hellmers, Herok, Hillmich, Horii, Howington, Hu, Hu, Huang,
  Huisman, Imai, Imamichi, Ishizaki, Ishwor, Iten, Itoko, Javadi,
  Javadi-Abhari, Javed, Jivrajani, Johns, Johnstun, Jonathan-Shoemaker,
  JosDenmark, JoshDumo, Judge, Kachmann, Kale, Kanazawa, Kane, Kang-Bae,
  Kapila, Karazeev, Kassebaum, Kelso, Kelso, Khanderao, King, Kobayashi,
  Kovyrshin, Krishnakumar, Krishnan, Krsulich, Kumkar, Kus, LaRose, Lacal,
  Lambert, Lapeyre, Latone, Lawrence, Lee, Li, Lishman, Liu, Liu, Maeng,
  Maheshkar, Majmudar, Malyshev, Mandouh, Manela, Manjula, Marecek, Marques,
  Marwaha, Maslov, Maszota, Mathews, Matsuo, Mazhandu, McClure, McElaney,
  McGarry, McKay, McPherson, Meesala, Meirom, Mendell, Metcalfe, Mevissen,
  Meyer, Mezzacapo, Midha, Minev, Mitchell, Moll, Montanez, Monteiro, Mooring,
  Morales, Moran, Morcuende, Mostafa, Motta, Moyard, Murali, M{\"u}ggenburg,
  Nadlinger, Nakanishi, Nannicini, Nation, Navarro, Naveh, Neagle, Neuweiler,
  Ngoueya, Nicander, Nick-Singstock, Niroula, Norlen, NuoWenLei, O'Riordan,
  Ogunbayo, Ollitrault, Onodera, Otaolea, Oud, Padilha, Paik, Pal, Pang,
  Panigrahi, Pascuzzi, Perriello, Peterson, Phan, Piro, Pistoia, Piveteau,
  Plewa, Pocreau, Pozas-Kerstjens, Pracht, Prokop, Prutyanov, Puri, Puzzuoli,
  P{\'e}rez, Quintiii, Rahman, Raja, Rajeev, Ramagiri, Rao, Raymond,
  Reardon-Smith, Redondo, Reuter, Rice, Riedemann, Risinger, Rocca,
  Rodr{\'\i}guez, RohithKarur, Rosand, Rossmannek, Ryu, SAPV, Saha, Ash-Saki,
  Sandberg, Sandesara, Sapra, Sargsyan, Sarkar, Sathaye, Schmitt, Schnabel,
  Schoenfeld, Scholten, Schoute, Schulterbrandt, Schwarm, Seaward, Sergi,
  Sertage, Setia, Shah, Shammah, Sharma, Shi, Shoemaker, Silva, Simonetto,
  Singh, Singh, Singkanipa, Siraichi, Siri, Sistos, Sitdikov, Sivarajah,
  Sletfjerding, Smolin, Soeken, Sokolov, Sokolov, SooluThomas, Starfish,
  Steenken, Stypulkoski, Suau, Sun, Sung, Suwama, S{\l}owik, Takahashi,
  Takawale, Tavernelli, Taylor, Taylour, Thomas, Tillet, Tod, Tomasik, de~la
  Torre, Toural, Trabing, Treinish, Trenev, TrishaPe, Truger, Tsilimigkounakis,
  Tulsi, Turner, Vaknin, Valcarce, Varchon, Vartak, Vazquez, Vijaywargiya,
  Villar, Vishnu, Vogt-Lee, Vuillot, Weaver, Weidenfeller, Wieczorek,
  Wildstrom, Wilson, Winston, WinterSoldier, Woehr, Woerner, Woo, Wood, Wood,
  Wood, Wootton, Wright, Yang, Yeralin, Yonekura, Yonge-Mallo, Young, Yu, Yu,
  Zachow, Zdanski, Zhang, Zoufal, {aeddins-ibm}, alexzhang13, b63,
  {bartek-bartlomiej}, bcamorrison, brandhsn, catornow, charmerDark,
  deeplokhande, dekel.meirom, dime10, ehchen, fanizzamarco, fs1132429, gadial,
  galeinston, georgezhou20, {georgios-ts}, gruu, hhorii, hykavitha, itoko,
  jliu45, jscott2, klinvill, krutik2966, ma5x, michelle4654, msuwama, ntgiwsvp,
  ordmoj, sagar pahwa, pritamsinha2304, ryancocuzzo, {saswati-qiskit},
  septembrr, sethmerkel, shaashwat, sternparky, strickroman, tigerjack,
  {tsura-crisaldo}, welien, willhbang, yang.luh, and
  {\v{C}}epulkovskis]{Qiskit}
MD~SAJID ANIS, H{\'e}ctor Abraham, AduOffei, Rochisha Agarwal, Gabriele
  Agliardi, Merav Aharoni, Ismail~Yunus Akhalwaya, Gadi Aleksandrowicz, Thomas
  Alexander, Matthew Amy, Sashwat Anagolum, Eli Arbel, Abraham Asfaw, Anish
  Athalye, Artur Avkhadiev, Carlos Azaustre, Abhik Banerjee, Santanu Banerjee,
  Will Bang, Aman Bansal, Panagiotis Barkoutsos, Ashish Barnawal, George
  Barron, George~S. Barron, Luciano Bello, Yael Ben-Haim, Daniel Bevenius,
  Dhruv Bhatnagar, Arjun Bhobe, Paolo Bianchini, Lev~S. Bishop, Carsten Blank,
  Sorin Bolos, Soham Bopardikar, Samuel Bosch, Sebastian Brandhofer, Brandon,
  Sergey Bravyi, Nick Bronn, Bryce-Fuller, David Bucher, Artemiy Burov, Fran
  Cabrera, Padraic Calpin, Lauren Capelluto, Jorge Carballo, Gin{\'e}s
  Carrascal, Adam Carriker, Ivan Carvalho, Adrian Chen, Chun-Fu Chen, Edward
  Chen, Jielun~(Chris) Chen, Richard Chen, Franck Chevallier, Rathish
  Cholarajan, Jerry~M. Chow, Spencer Churchill, Christian Claus, Christian
  Clauss, Caleb Clothier, Romilly Cocking, Ryan Cocuzzo, Jordan Connor, Filipe
  Correa, Abigail~J. Cross, Andrew~W. Cross, Simon Cross, Juan Cruz-Benito,
  Chris Culver, Antonio~D. C{\'o}rcoles-Gonzales, Navaneeth D, Sean Dague,
  Tareq~El Dandachi, Animesh~N Dangwal, Jonathan Daniel, Marcus Daniels,
  Matthieu Dartiailh, Abd{\'o}n~Rodr{\'\i}guez Davila, Faisal Debouni, Anton
  Dekusar, Amol Deshmukh, Mohit Deshpande, Delton Ding, Jun Doi, Eli~M. Dow,
  Eric Drechsler, Eugene Dumitrescu, Karel Dumon, Ivan Duran, Kareem EL-Safty,
  Eric Eastman, Grant Eberle, Amir Ebrahimi, Pieter Eendebak, Daniel Egger,
  Alberto Espiricueta, Mark Everitt, Davide Facoetti, Farida, Paco~Mart{\'\i}n
  Fern{\'a}ndez, Samuele Ferracin, Davide Ferrari, Axel~Hern{\'a}ndez Ferrera,
  Romain Fouilland, Albert Frisch, Andreas Fuhrer, Bryce Fuller, MELVIN GEORGE,
  Julien Gacon, Borja~Godoy Gago, Claudio Gambella, Jay~M. Gambetta, Adhisha
  Gammanpila, Luis Garcia, Tanya Garg, Shelly Garion, Tim Gates, Leron Gil,
  Austin Gilliam, Aditya Giridharan, Juan Gomez-Mosquera, Gonzalo, Salvador
  de~la Puente~Gonz{\'a}lez, Jesse Gorzinski, Ian Gould, Donny Greenberg,
  Dmitry Grinko, Wen Guan, John~A. Gunnels, Naman Gupta, Jakob~M. G{\"u}nther,
  Mikael Haglund, Isabel Haide, Ikko Hamamura, Omar~Costa Hamido, Frank
  Harkins, Areeq Hasan, Vojtech Havlicek, Joe Hellmers, {\L}ukasz Herok, Stefan
  Hillmich, Hiroshi Horii, Connor Howington, Shaohan Hu, Wei Hu, Junye Huang,
  Rolf Huisman, Haruki Imai, Takashi Imamichi, Kazuaki Ishizaki, Ishwor, Raban
  Iten, Toshinari Itoko, Ali Javadi, Ali Javadi-Abhari, Wahaj Javed, Madhav
  Jivrajani, Kiran Johns, Scott Johnstun, Jonathan-Shoemaker, JosDenmark,
  JoshDumo, John Judge, Tal Kachmann, Akshay Kale, Naoki Kanazawa, Jessica
  Kane, Kang-Bae, Annanay Kapila, Anton Karazeev, Paul Kassebaum, Josh Kelso,
  Scott Kelso, Vismai Khanderao, Spencer King, Yuri Kobayashi, Arseny
  Kovyrshin, Rajiv Krishnakumar, Vivek Krishnan, Kevin Krsulich, Prasad Kumkar,
  Gawel Kus, Ryan LaRose, Enrique Lacal, Rapha{\"e}l Lambert, John Lapeyre, Joe
  Latone, Scott Lawrence, Christina Lee, Gushu Li, Jake Lishman, Dennis Liu,
  Peng Liu, Yunho Maeng, Saurav Maheshkar, Kahan Majmudar, Aleksei Malyshev,
  Mohamed~El Mandouh, Joshua Manela, Manjula, Jakub Marecek, Manoel Marques,
  Kunal Marwaha, Dmitri Maslov, Pawe{\l} Maszota, Dolph Mathews, Atsushi
  Matsuo, Farai Mazhandu, Doug McClure, Maureen McElaney, Cameron McGarry,
  David McKay, Dan McPherson, Srujan Meesala, Dekel Meirom, Corey Mendell,
  Thomas Metcalfe, Martin Mevissen, Andrew Meyer, Antonio Mezzacapo, Rohit
  Midha, Zlatko Minev, Abby Mitchell, Nikolaj Moll, Alejandro Montanez, Gabriel
  Monteiro, Michael~Duane Mooring, Renier Morales, Niall Moran, David
  Morcuende, Seif Mostafa, Mario Motta, Romain Moyard, Prakash Murali, Jan
  M{\"u}ggenburg, David Nadlinger, Ken Nakanishi, Giacomo Nannicini, Paul
  Nation, Edwin Navarro, Yehuda Naveh, Scott~Wyman Neagle, Patrick Neuweiler,
  Aziz Ngoueya, Johan Nicander, Nick-Singstock, Pradeep Niroula, Hassi Norlen,
  NuoWenLei, Lee~James O'Riordan, Oluwatobi Ogunbayo, Pauline Ollitrault,
  Tamiya Onodera, Raul Otaolea, Steven Oud, Dan Padilha, Hanhee Paik, Soham
  Pal, Yuchen Pang, Ashish Panigrahi, Vincent~R. Pascuzzi, Simone Perriello,
  Eric Peterson, Anna Phan, Francesco Piro, Marco Pistoia, Christophe Piveteau,
  Julia Plewa, Pierre Pocreau, Alejandro Pozas-Kerstjens, Rafa{\l} Pracht,
  Milos Prokop, Viktor Prutyanov, Sumit Puri, Daniel Puzzuoli, Jes{\'u}s
  P{\'e}rez, Quintiii, Rafey~Iqbal Rahman, Arun Raja, Roshan Rajeev, Nipun
  Ramagiri, Anirudh Rao, Rudy Raymond, Oliver Reardon-Smith, Rafael
  Mart{\'\i}n-Cuevas Redondo, Max Reuter, Julia Rice, Matt Riedemann, Drew
  Risinger, Marcello~La Rocca, Diego~M. Rodr{\'\i}guez, RohithKarur, Ben
  Rosand, Max Rossmannek, Mingi Ryu, Tharrmashastha SAPV, Arijit Saha, Abdullah
  Ash-Saki, Martin Sandberg, Hirmay Sandesara, Ritvik Sapra, Hayk Sargsyan,
  Aniruddha Sarkar, Ninad Sathaye, Bruno Schmitt, Chris Schnabel, Zachary
  Schoenfeld, Travis~L. Scholten, Eddie Schoute, Mark Schulterbrandt, Joachim
  Schwarm, James Seaward, Sergi, Ismael~Faro Sertage, Kanav Setia, Freya Shah,
  Nathan Shammah, Rohan Sharma, Yunong Shi, Jonathan Shoemaker, Adenilton
  Silva, Andrea Simonetto, Divyanshu Singh, Parmeet Singh, Phattharaporn
  Singkanipa, Yukio Siraichi, Siri, Jes{\'u}s Sistos, Iskandar Sitdikov, Seyon
  Sivarajah, Magnus~Berg Sletfjerding, John~A. Smolin, Mathias Soeken,
  Igor~Olegovich Sokolov, Igor Sokolov, SooluThomas, Starfish, Dominik
  Steenken, Matt Stypulkoski, Adrien Suau, Shaojun Sun, Kevin~J. Sung, Makoto
  Suwama, Oskar S{\l}owik, Hitomi Takahashi, Tanvesh Takawale, Ivano
  Tavernelli, Charles Taylor, Pete Taylour, Soolu Thomas, Mathieu Tillet, Maddy
  Tod, Miroslav Tomasik, Enrique de~la Torre, Juan Luis~S{\'a}nchez Toural,
  Kenso Trabing, Matthew Treinish, Dimitar Trenev, TrishaPe, Felix Truger,
  Georgios Tsilimigkounakis, Davindra Tulsi, Wes Turner, Yotam Vaknin,
  Carmen~Recio Valcarce, Francois Varchon, Adish Vartak, Almudena~Carrera
  Vazquez, Prajjwal Vijaywargiya, Victor Villar, Bhargav Vishnu, Desiree
  Vogt-Lee, Christophe Vuillot, James Weaver, Johannes Weidenfeller, Rafal
  Wieczorek, Jonathan~A. Wildstrom, Jessica Wilson, Erick Winston,
  WinterSoldier, Jack~J. Woehr, Stefan Woerner, Ryan Woo, Christopher~J. Wood,
  Ryan Wood, Steve Wood, James Wootton, Matt Wright, Bo~Yang, Daniyar Yeralin,
  Ryota Yonekura, David Yonge-Mallo, Richard Young, Jessie Yu, Lebin Yu,
  Christopher Zachow, Laura Zdanski, Helena Zhang, Christa Zoufal,
  {aeddins-ibm}, alexzhang13, b63, {bartek-bartlomiej}, bcamorrison, brandhsn,
  catornow, charmerDark, deeplokhande, dekel.meirom, dime10, ehchen,
  fanizzamarco, fs1132429, gadial, galeinston, georgezhou20, {georgios-ts},
  gruu, hhorii, hykavitha, itoko, jliu45, jscott2, klinvill, krutik2966, ma5x,
  michelle4654, msuwama, ntgiwsvp, ordmoj, sagar pahwa, pritamsinha2304,
  ryancocuzzo, {saswati-qiskit}, septembrr, sethmerkel, shaashwat, sternparky,
  strickroman, tigerjack, {tsura-crisaldo}, welien, willhbang, yang.luh, and
  Mantas {\v{C}}epulkovskis.
\newblock Qiskit: An open-source framework for quantum computing, 2021.
\newblock URL \url{https://doi.org/10.5281/zenodo.2573505}.

\bibitem[{Cirq developers}(2021)]{Cirq}
{Cirq developers}.
\newblock Cirq, 2021.
\newblock URL \url{https://doi.org/10.5281/zenodo.5182845}.

\bibitem[West et~al.(2010)West, Lidar, Fong, and Gyure]{dynamicdecoupling2010}
Jacob~R. West, Daniel~A. Lidar, Bryan~H. Fong, and Mark~F. Gyure.
\newblock High fidelity quantum gates via dynamical decoupling.
\newblock \emph{Physical Review Letters}, 105\penalty0 (23), 2010.
\newblock ISSN 1079-7114.
\newblock \doi{10.1103/physrevlett.105.230503}.
\newblock URL \url{http://dx.doi.org/10.1103/PhysRevLett.105.230503}.

\bibitem[Das et~al.(2021)Das, Tannu, Dangwal, and Qureshi]{das2021adapt}
Poulami Das, Swamit Tannu, Siddharth Dangwal, and Moinuddin Qureshi.
\newblock {ADAPT}: Mitigating idling errors in qubits via adaptive dynamical
  decoupling.
\newblock In \emph{MICRO-54: 54th Annual IEEE/ACM International Symposium on
  Microarchitecture}, pages 950--962, 2021.
\newblock URL \url{https://doi.org/10.1145/3466752.3480059}.

\bibitem[Sundaresan et~al.(2020)Sundaresan, Lauer, Pritchett, Magesan,
  Jurcevic, and Gambetta]{sundaresan2020reducing}
Neereja Sundaresan, Isaac Lauer, Emily Pritchett, Easwar Magesan, Petar
  Jurcevic, and Jay~M Gambetta.
\newblock Reducing unitary and spectator errors in cross resonance with
  optimized rotary echoes.
\newblock \emph{PRX Quantum}, 1\penalty0 (2):\penalty0 020318, 2020.
\newblock URL \url{https://link.aps.org/doi/10.1103/PRXQuantum.1.020318}.

\bibitem[Sheldon et~al.(2016)Sheldon, Magesan, Chow, and
  Gambetta]{sheldon2016procedure}
Sarah Sheldon, Easwar Magesan, Jerry~M Chow, and Jay~M Gambetta.
\newblock Procedure for systematically tuning up cross-talk in the
  cross-resonance gate.
\newblock \emph{Physical Review A}, 93\penalty0 (6):\penalty0 060302, 2016.
\newblock URL \url{https://link.aps.org/doi/10.1103/PhysRevA.93.060302}.

\bibitem[Venturelli et~al.(2018)Venturelli, Do, Rieffel, and
  Frank]{temporalplanners2018}
Davide Venturelli, Minh Do, Eleanor Rieffel, and Jeremy Frank.
\newblock Compiling quantum circuits to realistic hardware architectures using
  temporal planners.
\newblock \emph{Quantum Science and Technology}, 3\penalty0 (2):\penalty0
  025004, 2018.
\newblock ISSN 2058-9565.
\newblock \doi{10.1088/2058-9565/aaa331}.
\newblock URL \url{http://dx.doi.org/10.1088/2058-9565/aaa331}.

\bibitem[Lao et~al.(2021)Lao, van Someren, Ashraf, and Almudever]{lao2021}
Lingling Lao, Hans van Someren, Imran Ashraf, and Carmen~G. Almudever.
\newblock Timing and resource-aware mapping of quantum circuits to
  superconducting processors.
\newblock \emph{IEEE Transactions on Computer-Aided Design of Integrated
  Circuits and Systems}, page 1–1, 2021.
\newblock ISSN 1937-4151.
\newblock \doi{10.1109/tcad.2021.3057583}.
\newblock URL \url{http://dx.doi.org/10.1109/TCAD.2021.3057583}.

\bibitem[Li et~al.(2019)Li, Ding, and Xie]{li2019tackling}
Gushu Li, Yufei Ding, and Yuan Xie.
\newblock Tackling the qubit mapping problem for {NISQ}-era quantum devices.
\newblock In \emph{Proceedings of the Twenty-Fourth International Conference on
  Architectural Support for Programming Languages and Operating Systems}, pages
  1001--1014, 2019.
\newblock URL \url{https://doi.org/10.1145/3297858.3304023}.

\bibitem[Tan and Cong(2021)]{tan2021optimal}
Bochen Tan and Jason Cong.
\newblock Optimal qubit mapping with simultaneous gate absorption.
\newblock In \emph{{IEEE}/{ACM} International Conference On Computer Aided
  Design}. {IEEE}, 2021.
\newblock \doi{10.1109/iccad51958.2021.9643554}.
\newblock URL \url{https://doi.org/10.1109%2Ficcad51958.2021.9643554}.

\bibitem[Tannu and Qureshi(2019{\natexlab{a}})]{tannu2019ensemble}
Swamit~S Tannu and Moinuddin Qureshi.
\newblock Ensemble of diverse mappings: Improving reliability of quantum
  computers by orchestrating dissimilar mistakes.
\newblock In \emph{Proceedings of the 52nd Annual IEEE/ACM International
  Symposium on Microarchitecture}, pages 253--265, 2019{\natexlab{a}}.
\newblock URL \url{https://doi.org/10.1145/3352460.3358257}.

\bibitem[Patel et~al.(2020)Patel, Li, Roy, and Tiwari]{patel2020ureqa}
Tirthak Patel, Baolin Li, Rohan~Basu Roy, and Devesh Tiwari.
\newblock {UREQA}: Leveraging operation-aware error rates for effective quantum
  circuit mapping on {NISQ}-era quantum computers.
\newblock In \emph{USENIX Annual Technical Conference}, pages 705--711, 2020.
\newblock URL \url{https://www.usenix.org/conference/atc20/presentation/patel}.

\bibitem[Tannu and Qureshi(2019{\natexlab{b}})]{tannu2019not}
Swamit~S Tannu and Moinuddin~K Qureshi.
\newblock Not all qubits are created equal: A case for variability-aware
  policies for {NISQ}-era quantum computers.
\newblock In \emph{Proceedings of the Twenty-Fourth International Conference on
  Architectural Support for Programming Languages and Operating Systems}, pages
  987--999, 2019{\natexlab{b}}.
\newblock URL \url{https://doi.org/10.1145/3297858.3304007}.

\bibitem[Duckering et~al.(2021)Duckering, Baker, Litteken, and
  Chong]{Duckering2021}
Casey Duckering, Jonathan~M. Baker, Andrew Litteken, and Frederic~T. Chong.
\newblock Orchestrated trios: Compiling for efficient communication in quantum
  programs with 3-qubit gates.
\newblock In \emph{Proceedings of the 26th {ACM} International Conference on
  Architectural Support for Programming Languages and Operating Systems}.
  {ACM}, 2021.
\newblock \doi{10.1145/3445814.3446718}.
\newblock URL \url{https://doi.org/10.1145/3445814.3446718}.

\bibitem[Finigan et~al.(2018)Finigan, Cubeddu, Lively, Flick, and
  Narang]{finigan2018qubit}
Will Finigan, Michael Cubeddu, Thomas Lively, Johannes Flick, and Prineha
  Narang.
\newblock Qubit allocation for noisy intermediate-scale quantum computers,
  2018.
\newblock URL \url{https://arxiv.org/abs/1810.08291}.

\bibitem[Murali et~al.(2020)Murali, McKay, Martonosi, and
  Javadi-Abhari]{murali2020software}
Prakash Murali, David~C McKay, Margaret Martonosi, and Ali Javadi-Abhari.
\newblock Software mitigation of crosstalk on noisy intermediate-scale quantum
  computers.
\newblock In \emph{Proceedings of the Twenty-Fifth International Conference on
  Architectural Support for Programming Languages and Operating Systems}, pages
  1001--1016, 2020.
\newblock URL \url{https://doi.org/10.1145/3373376.3378477}.

\bibitem[Tannu and Qureshi(2019{\natexlab{c}})]{tannu2019mitigating}
Swamit~S Tannu and Moinuddin~K Qureshi.
\newblock Mitigating measurement errors in quantum computers by exploiting
  state-dependent bias.
\newblock In \emph{Proceedings of the 52nd Annual IEEE/ACM International
  Symposium on Microarchitecture}, pages 279--290, 2019{\natexlab{c}}.
\newblock URL \url{https://doi.org/10.1145/3352460.3358265}.

\bibitem[Gokhale et~al.(2021)Gokhale, Tomesh, Suchara, and
  Chong]{gokhale2021faster}
Pranav Gokhale, Teague Tomesh, Martin Suchara, and Frederic~T. Chong.
\newblock Faster and more reliable quantum swaps via native gates, 2021.
\newblock URL \url{https://arxiv.org/abs/2109.13199}.

\bibitem[Gokhale et~al.(2020{\natexlab{a}})Gokhale, Javadi-Abhari, Earnest,
  Shi, and Chong]{gokhale2020optimized}
Pranav Gokhale, Ali Javadi-Abhari, Nathan Earnest, Yunong Shi, and Frederic~T.
  Chong.
\newblock Optimized quantum compilation for near-term algorithms with
  {OpenPulse}.
\newblock Technical Report 2004.11205, arXiv, 2020{\natexlab{a}}.
\newblock URL \url{https://arxiv.org/abs/2004.11205}.

\bibitem[Satoh et~al.(2021)Satoh, Oomura, Sugawara, and
  Yamamoto]{oomura2021design}
Takahiko Satoh, Shun Oomura, Michihiko Sugawara, and Naoki Yamamoto.
\newblock Pulse-engineered controlled-{V} gate and its applications on
  superconducting quantum device.
\newblock Technical Report 2102.06117, arXiv, 2021.
\newblock URL \url{https://arxiv.org/abs/2102.06117}.

\bibitem[Toffoli(1980)]{ToffoliReversibleComputing}
Tommaso Toffoli.
\newblock Reversible computing.
\newblock In Jaco de~Bakker and Jan van Leeuwen, editors, \emph{Automata,
  Languages and Programming}, pages 632--644, Berlin, Heidelberg, 1980.
  Springer Berlin Heidelberg.
\newblock ISBN 978-3-540-39346-7.
\newblock URL \url{https://doi.org/10.21236/ada082021}.

\bibitem[Shende and Markov(2009)]{shende2008cnotcost}
V.V. Shende and I.L. Markov.
\newblock On the {CNOT}-cost of {TOFFOLI} gates.
\newblock \emph{Quantum Information and Computation}, 9\penalty0
  (5{\&}6):\penalty0 461--486, 2009.
\newblock \doi{10.26421/qic8.5-6-8}.
\newblock URL \url{https://doi.org/10.26421%2Fqic8.5-6-8}.

\bibitem[Shor(1997)]{Shor_1997}
Peter~W. Shor.
\newblock Polynomial-time algorithms for prime factorization and discrete
  logarithms on a quantum computer.
\newblock \emph{SIAM Journal on Computing}, 26\penalty0 (5):\penalty0
  1484–1509, 1997.
\newblock ISSN 1095-7111.
\newblock \doi{10.1137/s0097539795293172}.
\newblock URL \url{http://dx.doi.org/10.1137/S0097539795293172}.

\bibitem[Grover(1996)]{Grover1996}
Lov~K. Grover.
\newblock A fast quantum mechanical algorithm for database search.
\newblock In \emph{Proceedings of the {ACM} Symposium on Theory of Computing}.
  {ACM} Press, 1996.
\newblock \doi{10.1145/237814.237866}.
\newblock URL \url{https://doi.org/10.1145/237814.237866}.

\bibitem[Figgatt et~al.(2017)Figgatt, Maslov, Landsman, Linke, Debnath, and
  Monroe]{Figgatt2017}
C.~Figgatt, D.~Maslov, K.~A. Landsman, N.~M. Linke, S.~Debnath, and C.~Monroe.
\newblock Complete 3-qubit {Grover} search on a programmable quantum computer.
\newblock \emph{Nature Communications}, 8\penalty0 (1), 2017.
\newblock \doi{10.1038/s41467-017-01904-7}.
\newblock URL \url{https://doi.org/10.1038/s41467-017-01904-7}.

\bibitem[Takahashi et~al.(2010)Takahashi, Tani, and
  Kunihiro]{takahashi2009quantum}
Yasuhiro Takahashi, Seiichiro Tani, and Noboru Kunihiro.
\newblock Quantum addition circuits and unbounded fan-out.
\newblock \emph{Quantum Information and Computation}, 10\penalty0
  (9{\&}10):\penalty0 872--890, 2010.
\newblock \doi{10.26421/qic10.9-10-12}.
\newblock URL \url{https://doi.org/10.26421%2Fqic10.9-10-12}.

\bibitem[Reed et~al.(2012)Reed, DiCarlo, Nigg, Sun, Frunzio, Girvin, and
  Schoelkopf]{Reed_2012}
M.~D. Reed, L.~DiCarlo, S.~E. Nigg, L.~Sun, L.~Frunzio, S.~M. Girvin, and R.~J.
  Schoelkopf.
\newblock Realization of three-qubit quantum error correction with
  superconducting circuits.
\newblock \emph{Nature}, 482\penalty0 (7385):\penalty0 382–385, 2012.
\newblock ISSN 1476-4687.
\newblock \doi{10.1038/nature10786}.
\newblock URL \url{http://dx.doi.org/10.1038/nature10786}.

\bibitem[Crow et~al.(2016)Crow, Joynt, and Saffman]{crow2016improved}
Daniel Crow, Robert Joynt, and Mark Saffman.
\newblock Improved error thresholds for measurement-free error correction.
\newblock \emph{Physical Review Letters}, 117\penalty0 (13):\penalty0 130503,
  2016.
\newblock \doi{10.1103/PhysRevLett.117.130503}.

\bibitem[Hill and Marty(2008)]{Hill2008}
Mark~D. Hill and Michael~R. Marty.
\newblock Amdahl{\textquotesingle}s law in the multicore era.
\newblock \emph{Computer}, 41\penalty0 (7):\penalty0 33--38, 2008.
\newblock \doi{10.1109/mc.2008.209}.
\newblock URL \url{https://doi.org/10.1109/mc.2008.209}.

\bibitem[Cowtan et~al.(2019)Cowtan, Dilkes, Duncan, Krajenbrink, Simmons, and
  Sivarajah]{cowtan_et_al:LIPIcs:2019:10397}
Alexander Cowtan, Silas Dilkes, Ross Duncan, Alexandre Krajenbrink, Will
  Simmons, and Seyon Sivarajah.
\newblock {On the Qubit Routing Problem}.
\newblock In Wim van Dam and Laura Mancinska, editors, \emph{14th Conference on
  the Theory of Quantum Computation, Communication and Cryptography}, volume
  135 of \emph{Leibniz International Proceedings in Informatics (LIPIcs)},
  pages 5:1--5:32, Dagstuhl, Germany, 2019. Schloss Dagstuhl--Leibniz-Zentrum
  fuer Informatik.
\newblock ISBN 978-3-95977-112-2.
\newblock \doi{10.4230/LIPIcs.TQC.2019.5}.
\newblock URL \url{http://drops.dagstuhl.de/opus/volltexte/2019/10397}.

\bibitem[Gokhale et~al.(2020{\natexlab{b}})Gokhale, Koretsky, Huang, Majumder,
  Drucker, Brown, and Chong]{gokhale2020quantum}
Pranav Gokhale, Samantha Koretsky, Shilin Huang, Swarnadeep Majumder, Andrew
  Drucker, Kenneth~R. Brown, and Frederic~T. Chong.
\newblock Quantum fan-out: Circuit optimizations and technology modeling.
\newblock Technical Report 2007.04246, arXiv, 2020{\natexlab{b}}.
\newblock URL \url{https://arxiv.org/abs/2007.04246}.

\bibitem[Nation et~al.(2021)Nation, Paik, Cross, and Nazario]{Ibm2021}
Paul Nation, Hanhee Paik, Andrew Cross, and Zaira Nazario.
\newblock The {IBM} {Quantum} heavy hex lattice.
\newblock Technical report, IBM, 2021.
\newblock URL \url{https://www.research.ibm.com/blog/heavy-hex-lattice}.

\bibitem[Nielsen and Chuang(2009)]{Nielsen2009}
Michael~A. Nielsen and Isaac~L. Chuang.
\newblock \emph{Quantum Computation and Quantum Information}.
\newblock Cambridge University Press, 2009.
\newblock \doi{10.1017/cbo9780511976667}.
\newblock URL \url{https://doi.org/10.1017/cbo9780511976667}.

\bibitem[{SuperstaQ Development Team}(2021)]{superstaq}
{SuperstaQ Development Team}.
\newblock {SuperstaQ}: Connecting applications to quantum hardware.
\newblock \url{www.super.tech/about-superstaq}, 2021.

\bibitem[McKay et~al.(2017)McKay, Wood, Sheldon, Chow, and
  Gambetta]{mckay2017efficient}
David~C McKay, Christopher~J Wood, Sarah Sheldon, Jerry~M Chow, and Jay~M
  Gambetta.
\newblock Efficient z gates for quantum computing.
\newblock \emph{Physical Review A}, 96\penalty0 (2):\penalty0 022330, 2017.

\bibitem[Cheng et~al.(2020)Cheng, Deng, and Qian]{cheng2020accqoc}
Jinglei Cheng, Haoqing Deng, and Xuehai Qian.
\newblock {AccQOC}: Accelerating quantum optimal control based pulse
  generation, 2020.
\newblock URL \url{https://doi.org/10.1109/ISCA45697.2020.00052}.

\bibitem[Barends et~al.(2014)Barends, Kelly, Megrant, Veitia, Sank, Jeffrey,
  White, Mutus, Fowler, Campbell, Chen, Chen, Chiaro, Dunsworth, Neill,
  O'Malley, Roushan, Vainsencher, Wenner, Korotkov, Cleland, and
  Martinis]{barends2014superconductingmontecarlo}
R.~Barends, J.~Kelly, A.~Megrant, A.~Veitia, D.~Sank, E.~Jeffrey, T.~C. White,
  J.~Mutus, A.~G. Fowler, B.~Campbell, Y.~Chen, Z.~Chen, B.~Chiaro,
  A.~Dunsworth, C.~Neill, P.~O'Malley, P.~Roushan, A.~Vainsencher, J.~Wenner,
  A.~N. Korotkov, A.~N. Cleland, and John~M. Martinis.
\newblock Superconducting quantum circuits at the surface code threshold for
  fault tolerance.
\newblock \emph{Nature}, 508\penalty0 (7497):\penalty0 500--503, 2014.
\newblock \doi{10.1038/nature13171}.
\newblock URL \url{http://dx.doi.org/10.1038/nature13171}.

\bibitem[Huang et~al.(2019)Huang, Yang, Chan, Tanttu, Hensen, Leon, Fogarty,
  Hwang, Hudson, Itoh, Morello, Laucht, and
  Dzurak]{huang2019fidelitymontecarlo2}
W.~Huang, C.~H. Yang, K.~W. Chan, T.~Tanttu, B.~Hensen, R.~C.~C. Leon, M.~A.
  Fogarty, J.~C.~C. Hwang, F.~E. Hudson, K.~M. Itoh, A.~Morello, A.~Laucht, and
  A.~S. Dzurak.
\newblock Fidelity benchmarks for two-qubit gates in silicon.
\newblock \emph{Nature}, 569\penalty0 (7757):\penalty0 532--536, 2019.
\newblock URL \url{https://doi.org/10.1038/s41586-019-1197-0}.

\bibitem[Noiri et~al.(2022)Noiri, Takeda, Nakajima, Kobayashi, Sammak,
  Scappucci, and Tarucha]{noiri2022fastmontecarlo3}
Akito Noiri, Kenta Takeda, Takashi Nakajima, Takashi Kobayashi, Amir Sammak,
  Giordano Scappucci, and Seigo Tarucha.
\newblock Fast universal quantum gate above the fault-tolerance threshold in
  silicon.
\newblock \emph{Nature}, 601\penalty0 (7893):\penalty0 338--342, 2022.
\newblock URL \url{https://doi.org/10.1038/s41586-021-04182-y}.

\bibitem[Smith et~al.(2021)Smith, Davis, Larson, Younis, Iancu, and
  Lavrijsen]{resynthesis2021}
Ethan Smith, Marc~G. Davis, Jeffrey Larson, Ed~Younis, Costin Iancu, and Wim
  Lavrijsen.
\newblock Leap: Scaling numerical optimization based synthesis using an
  incremental approach, 2021.
\newblock URL \url{https://arxiv.org/abs/2106.11246}.

\bibitem[Gokhale et~al.(2019)Gokhale, Ding, Propson, Winkler, Leung, Shi,
  Schuster, Hoffmann, and Chong]{Gokhale_2019}
Pranav Gokhale, Yongshan Ding, Thomas Propson, Christopher Winkler, Nelson
  Leung, Yunong Shi, David~I. Schuster, Henry Hoffmann, and Frederic~T. Chong.
\newblock Partial compilation of variational algorithms for noisy
  intermediate-scale quantum machines.
\newblock \emph{Proceedings of the 52nd Annual IEEE/ACM International Symposium
  on Microarchitecture}, 2019.
\newblock \doi{10.1145/3352460.3358313}.
\newblock URL \url{http://dx.doi.org/10.1145/3352460.3358313}.

\bibitem[Leung et~al.(2017)Leung, Abdelhafez, Koch, and Schuster]{Leung_2017}
Nelson Leung, Mohamed Abdelhafez, Jens Koch, and David Schuster.
\newblock Speedup for quantum optimal control from automatic differentiation
  based on graphics processing units.
\newblock \emph{Physical Review A}, 95\penalty0 (4), 2017.
\newblock ISSN 2469-9934.
\newblock \doi{10.1103/physreva.95.042318}.
\newblock URL \url{http://dx.doi.org/10.1103/PhysRevA.95.042318}.

\bibitem[Mor{\'e}(1978)]{more1978levenberg}
Jorge~J Mor{\'e}.
\newblock The {Levenberg}-{Marquardt} algorithm: Implementation and theory.
\newblock In \emph{Numerical Analysis}, pages 105--116. Springer, 1978.

\bibitem[Team(2021)]{QuOpt}
QuOpt~Development Team.
\newblock Quopt research software.
\newblock \url{https://github.com/maxaksel/QuOpt}, 2021.

\bibitem[Agarwal et~al.(2022)Agarwal, Mierle, and {The Ceres Solver
  Team}]{ceres-solver}
Sameer Agarwal, Keir Mierle, and {The Ceres Solver Team}.
\newblock {Ceres Solver}, 3 2022.
\newblock URL \url{https://github.com/ceres-solver/ceres-solver}.

\bibitem[Virtanen et~al.(2020)Virtanen, Gommers, Oliphant, Haberland, Reddy,
  Cournapeau, Burovski, Peterson, Weckesser, Bright, {van der Walt}, Brett,
  Wilson, Millman, Mayorov, Nelson, Jones, Kern, Larson, Carey, Polat, Feng,
  Moore, {VanderPlas}, Laxalde, Perktold, Cimrman, Henriksen, Quintero, Harris,
  Archibald, Ribeiro, Pedregosa, {van Mulbregt}, and {SciPy 1.0
  Contributors}]{2020SciPy-NMeth}
Pauli Virtanen, Ralf Gommers, Travis~E. Oliphant, Matt Haberland, Tyler Reddy,
  David Cournapeau, Evgeni Burovski, Pearu Peterson, Warren Weckesser, Jonathan
  Bright, St{\'e}fan~J. {van der Walt}, Matthew Brett, Joshua Wilson, K.~Jarrod
  Millman, Nikolay Mayorov, Andrew R.~J. Nelson, Eric Jones, Robert Kern, Eric
  Larson, C~J Carey, {\.I}lhan Polat, Yu~Feng, Eric~W. Moore, Jake
  {VanderPlas}, Denis Laxalde, Josef Perktold, Robert Cimrman, Ian Henriksen,
  E.~A. Quintero, Charles~R. Harris, Anne~M. Archibald, Ant{\^o}nio~H. Ribeiro,
  Fabian Pedregosa, Paul {van Mulbregt}, and {SciPy 1.0 Contributors}.
\newblock {{SciPy} 1.0: Fundamental Algorithms for Scientific Computing in
  Python}.
\newblock \emph{Nature Methods}, 17:\penalty0 261--272, 2020.
\newblock \doi{10.1038/s41592-019-0686-2}.
\newblock URL \url{https://doi.org/10.1038%2Fs41592-019-0686-2}.

\bibitem[Team(2022)]{PyQuOpt}
PyQuOpt~Development Team.
\newblock Pyquopt research software.
\newblock \url{https://github.com/maxaksel/Toffoli-Optimization}, 2022.

\bibitem[Baker et~al.(2020)Baker, Duckering, Gokhale, and
  Litteken]{BakerBenchmarks}
Jonathan~M. Baker, Casey Duckering, Pranav Gokhale, and Andrew Litteken.
\newblock Quantum circuit benchmarks.
\newblock \url{https://github.com/jmbaker94/quantumcircuitbenchmarks}, 2020.

\bibitem[Tomesh et~al.(2022)Tomesh, Allen, and Saleem]{tomesh2022qaoa2}
Teague Tomesh, Nicholas Allen, and Zain Saleem.
\newblock Quantum-classical tradeoffs and multi-controlled quantum gate
  decompositions in variational algorithms.
\newblock \emph{arXiv:2210.04378}, 2022.
\newblock \doi{10.48550/arXiv.2210.04378}.

\bibitem[Liu et~al.(2023)Liu, Bowman, Gokhale, Dangwal, Larson, Chong, and
  Hovland]{QContext}
Ji~Liu, Max Bowman, Pranav Gokhale, Siddharth Dangwal, Jeffrey Larson,
  Frederic~T. Chong, and Paul~D. Hovland.
\newblock Qcontext: Context-aware decomposition for quantum gates, 2023.
\newblock URL \url{https://arxiv.org/abs/2302.02003}.

\bibitem[Baker et~al.(2019)Baker, Duckering, Hoover, and
  Chong]{baker2019decomposing}
Jonathan~M. Baker, Casey Duckering, Alexander Hoover, and Frederic~T. Chong.
\newblock Decomposing quantum generalized {Toffoli} with an arbitrary number of
  ancilla, 2019.
\newblock URL \url{https://arxiv.org/abs/1904.01671}.

\bibitem[Gidney(2015)]{gidney_2015}
Craig Gidney.
\newblock Constructing large controlled nots, 2015.
\newblock URL
  \url{https://algassert.com/circuits/2015/06/05/Constructing-Large-Controlled-Nots.html}.

\bibitem[Barenco et~al.(1995)Barenco, Bennett, Cleve, DiVincenzo, Margolus,
  Shor, Sleator, Smolin, and Weinfurter]{Barenco_1995}
Adriano Barenco, Charles~H. Bennett, Richard Cleve, David~P. DiVincenzo, Norman
  Margolus, Peter Shor, Tycho Sleator, John~A. Smolin, and Harald Weinfurter.
\newblock Elementary gates for quantum computation.
\newblock \emph{Physical Review A}, 52\penalty0 (5):\penalty0 3457–3467,
  1995.
\newblock ISSN 1094-1622.
\newblock \doi{10.1103/physreva.52.3457}.
\newblock URL \url{http://dx.doi.org/10.1103/PhysRevA.52.3457}.

\bibitem[Cuccaro et~al.(2004)Cuccaro, Draper, Kutin, and
  Moulton]{quant-ph/0410184}
Steven~A. Cuccaro, Thomas~G. Draper, Samuel~A. Kutin, and David~Petrie Moulton.
\newblock A new quantum ripple-carry addition circuit, 2004.
\newblock URL \url{https://arxiv.org/abs/quant-ph/0410184}.

\bibitem[Hadfield et~al.(2019)Hadfield, Wang, O'Gorman, Rieffel, Venturelli,
  and Biswas]{hadfield2019qaoa1}
Stuart Hadfield, Zhihui Wang, Bryan O'Gorman, Eleanor~G Rieffel, Davide
  Venturelli, and Rupak Biswas.
\newblock From the quantum approximate optimization algorithm to a quantum
  alternating operator ansatz.
\newblock \emph{Algorithms}, 12\penalty0 (2):\penalty0 34, 2019.
\newblock \doi{10.3390/a12020034}.

\bibitem[Zhang et~al.(2022)Zhang, Majumder, Leung, Crain, Wang, Fang, Debroy,
  Kim, and Brown]{zhang2022hidden}
Bichen Zhang, Swarnadeep Majumder, Pak~Hong Leung, Stephen Crain, Ye~Wang, Chao
  Fang, Dripto~M Debroy, Jungsang Kim, and Kenneth~R Brown.
\newblock Hidden inverses: {C}oherent error cancellation at the circuit level.
\newblock \emph{Physical Review Applied}, 17\penalty0 (3):\penalty0 034074,
  2022.
\newblock \doi{10.1103/PhysRevApplied.17.034074}.

\bibitem[Levine et~al.(2019)Levine, Keesling, Semeghini, Omran, Wang, Ebadi,
  Bernien, Greiner, Vuleti{\'c}, Pichler, et~al.]{levine2019parallel}
Harry Levine, Alexander Keesling, Giulia Semeghini, Ahmed Omran, Tout~T Wang,
  Sepehr Ebadi, Hannes Bernien, Markus Greiner, Vladan Vuleti{\'c}, Hannes
  Pichler, et~al.
\newblock Parallel implementation of high-fidelity multiqubit gates with
  neutral atoms.
\newblock \emph{Physical review letters}, 123\penalty0 (17):\penalty0 170503,
  2019.
\newblock \doi{10.1103/PhysRevLett.123.170503}.

\end{thebibliography}

\end{document}